\documentclass[11pt,oneside]{thesis}
\usepackage{epsf}
\usepackage{stops}

\begin{document}
\let\footnotesize\normalsize
\addtolength{\baselineskip}{2ex}
\renewcommand{\thepage}{\roman{page}}
\thispagestyle{empty}
\addtolength{\baselineskip}{-2ex}
{\Large 
\vskip 10mm
\begin{center}
TECHNIQUES FOR THE \\
TOP SQUARK SEARCH AT THE FERMILAB TEVATRON
\end{center}
}
\vskip 40mm
\begin{center}
A DISSERTATION SUBMITTED TO THE GRADUATE DIVISION OF THE \\
UNIVERSITY OF HAWAI`I IN PARTIAL FULFILLMENT OF THE \\
REQUIREMENTS FOR THE DEGREE OF \\
\vskip 2.5ex
DOCTOR OF PHILOSOPHY \\
\vskip 2.5ex
IN \\
\vskip 2.5ex
PHYSICS \\
\vskip 6ex
AUGUST 2000
\vskip 20ex
By \\
\vskip .5ex
{\large John Sender}
\vskip 6ex
Dissertation Committee:
\vskip 2.5ex
Xerxes Tata, Chairperson\\
Sandip Pakvasa\\
Michael Peters\\
Chester Vause\\
Robert Joseph
\end{center}
\addtolength{\baselineskip}{2ex}

\newpage
We certify that we have read this dissertation and that,
in our opinion, it is satisfactory in scope and quality
as a dissertation for the degree of
Doctor of Philosophy in Physics.
\vskip 2.5in

\vskip 0.2in\hspace{2.6in}DISSERTATION COMMITTEE

\vskip 0.46in\hspace{2.2in}\hrulefill
\vskip -0.15in           \hspace{3.3in} Chairperson

\vskip 0.23in\hspace{2.2in}\hrulefill

\vskip 0.33in\hspace{2.2in}\hrulefill

\vskip 0.33in\hspace{2.2in}\hrulefill

\vskip 0.33in\hspace{2.2in}\hrulefill

%
%

\Abstract\addcontentsline{toc}{chapter}{Abstract}
This dissertation addresses the question of how to detect light top
squarks at the upgraded Fermilab Tevatron collider.  After a brief
introduction to supersymmetry, the basic phenomenology of the light
stop is reviewed and the current experimental situation is surveyed.
The analysis presented here is based on collider event simulations.
The main decay modes accessible to the Tevatron are studied, feasible
discovery channels are identified, and recipes for experimental
analysis are proposed.  It is found that stops with masses up to the
top quark mass are liable to detection under these schemes with the
data from a few years' running at the upgraded Tevatron.  With such an
extended run, significant portions of parameter space may be probed.

\def\contentsname{Table of Contents}
\addtolength{\baselineskip}{1ex}
\tableofcontents
\addtolength{\baselineskip}{-1ex}
\clearpage\addcontentsline{toc}{chapter}{List of Tables}\listoftables
\clearpage\addcontentsline{toc}{chapter}{List of Figures}\listoffigures
\clearpage

\renewcommand{\thepage}{\arabic{page}}
\setcounter{page}1
\widowpenalty 150
\clubpenalty 150
\setlength{\parindent}{3em}

\chapter{Introduction to Supersymmetry%
\label{chap:intro}%
}

This thesis investigates strategies for detecting a top squark (also
referred to as the scalar top or, for short, stop) at the Fermilab Tevatron $%
p\bar{p}$ collider. The top squark is a representative of a new class of
particles in the theory of supersymmetry. It is thought by some that the
stop could be one of the most accessible particles of this class, and
experimental physicists have invested a great deal of effort in the search
for this particle.

We begin in this Chapter with an introduction to supersymmetry and a brief
exposition of its attractions for particle physicists. We characterize the
class of practical supersymmetry models and offer a synopsis of the
phenomenology relevant to our study. Then we introduce the top squark and
discuss its properties in Chapter \ref{chap:stop}. We present the basic
formulae and indicate how the particle may be light, and also why it might
be expected to be light. We summarize the production and decay modes of the
stop, and include a \textit{pr\'{e}cis} of current experimental results on
the stop and other supersymmetric particles important to our work. In
Chapter \ref{chap:tech}, we describe calculational techniques and the
methodology of our analysis.

Our analysis of how experiments at the Tevatron may search for the stop
begins in Chapter \ref{chap:scz} with an investigation of the decay mode ${%
\tilde{t}_{1}}\rightarrow c{\widetilde{Z}_{1}}$. We review the event
topology and identify physics background processes which can mask its
signal. We look in several channels, and by examining key kinematical
features of the stop events we are led to propose a program of experimental
procedures designed to maximize the discovery potential of the stop decays
via this mode. We perform a similar study for the decay mode ${\tilde{t}_{1}}%
\rightarrow b{\widetilde{W}_{1}}$ in Chapter \ref{chap:sbw}. Our conclusions
and some summary remarks on the prospects for uncovering the stop at the
Tevatron are presented in Chapter \ref{chap:summ}.

\section{Supersymmetry as a fundamental symmetry of nature}

The body of data presently available from investigations into the
fundamental constituents of matter is consistent with a theoretical
framework known as the Standard Model (SM) of particle physics. All observed
particles and forces (apart from gravity) are encompassed by this theory.
The SM is a relativistic quantum field theory which organizes elementary
particles and the forces which act upon them according to the principles of
gauge symmetry\cite{Weinberg}. Despite its success, though, the Standard
Model has not furnished a satisfyingly complete explanation of the realm of
elementary particles. There are many \textit{ad hoc} parameters, and
difficulties arise when the theory is extrapolated to energies much beyond a
TeV (in the sense explained below) without some modification. The Standard
Model also affords no link between the gauge forces and the force of gravity.

Supersymmetry (SUSY)\cite{Nilles,NathAC,HaberKane}\ is a proposed new
symmetry of particle interactions which generalizes the symmetries of
space-time. The usual space-time symmetries are encoded in a set of
commutation relations for the generators of translation $P_{m}$ and rotation 
$M_{mn}$\cite{wignerPoin}: 
\begin{eqnarray}
\begin{array}{c}
\left[ M_{mn},M_{pq}\right] =_{\mathstrut _{\mathstrut _{\mathstrut
}}}g_{np}M_{mq}-g_{mp}M_{nq}-g_{nq}M_{mp}+g_{mq}M_{np}, \\ 
\left[ M_{mn},P_{q}\right] =^{\mathstrut ^{\mathstrut ^{\mathstrut
}}}g_{nq}P_{m}-g_{mq}P_{n},\quad \left[ P_{m},P_{n}\right] =0,
\end{array}
\label{eq:poincare}%
\end{eqnarray}
with $g_{mn}$ the metric tensor. These commutation relations constitute a
Lie algebra for the operators, called the Poincar\'{e} algebra. With
reasonable assumptions, Coleman and Mandula\cite{ColeMand} showed that any
Lie group that contains both the Lie group underlying the Poincar\'{e}
algebra (the Poincar\'{e} group) and an internal symmetry group must be just
a direct product of the Poincar\'{e} group and that symmetry group. There is
thus no non-trivial extension of the Poincar\'{e} group, and one can say
that the Poincar\'{e} algebra is the most general Lie algebra of space-time
symmetries.

The notion of a Lie algebra can be extended to that of a \emph{graded} Lie
algebra by admitting anticommutators in addition to commutators. The SUSY
algebra comes from adding so-called ``supersymmetry'' generators $Q^{a}$ to
the Poincar\'{e} algebra (\ref{eq:poincare}) satisfying the
(anti)commutation relations 
\begin{eqnarray}
\begin{array}{c}
\left\{ Q^{a},\bar{Q}^{b}\right\} =_{\mathstrut _{\mathstrut _{\mathstrut
}}}\left( \gamma ^{m}\right) ^{ab}P_{m},\quad \left\{ Q^{a},Q^{b}\right\} =0
\\ 
\left[ P_{m},Q^{a}\right] =^{\mathstrut ^{\mathstrut ^{\mathstrut }}}0,\quad 
\left[ M_{mn},Q^{a}\right] =\frac{1}{2}\left( \sigma _{mn}Q\right) ^{a},
\end{array}
\label{eq:susyalg}%
\end{eqnarray}
where $\gamma $ and $\sigma $ are the usual elements of the Dirac calculus.
Haag, \L opuszanski and Sohnius\cite{HLandS} extended the work of Coleman
and Mandula to show that the SUSY algebra of equations (\ref{eq:poincare})
and (\ref{eq:susyalg}) constitutes the most general graded Lie algebra of
space-time symmetries.

The SUSY generators in (\ref{eq:susyalg}) are spinorial. Spinors, though
non-classical, are indispensable in field theories of matter as we know it.
They furnish the representation for electrons (and all matter particles),
but under the algebra (\ref{eq:poincare}) they are contingent objects in
space-time. SUSY builds spinors into the structure of space-time itself and
so confers on them an ontological status more befitting their fundamental
role. Metaphysically, this is tidier.

In the usual SUSY scheme the transformations of equation \ref{eq:susyalg}
are global. If we promote these to a \emph{local} symmetry then remarkably
we derive the general coordinate transformations of general relativity, and
find spin-2 gravitons in the theory\cite{localSUSY1}\cite{localSUSY2}. Local
supersymmetry necessarily leads to quantum gravity. No one has yet been able
to write down a consistent field theory of quantum gravity to capitalize on
this fact, but it is intriguing and suggests that SUSY may have a basic role
to play in the unification of the gauge forces with gravity. The glamorous
``superstring'' theory, which \emph{is} a finite theory of quantum gravity,
is founded on space-time supersymmetry.

The notion of unification is on a par with symmetry as a fundamental
organizational principle in physics, as exemplified by\ Newton's
identification of terrestrial and planetary gravity, Maxwell's synthesis of
electricity and magnetism into a single field, or Einstein's equivalence of
gravitational and inertial mass. In particle physics we can point to the
unification\footnote{%
Many do not count GSW theory as a proper unification, since one needs as
many parameters with the theory as without; it is, however, a satisfying 
\emph{conceptual} unification that puts the electromagnetic and weak forces
on a common footing and serves as a model for GUT unification.} of the
electromagnetic and weak forces in Glashow, Salam and Weinberg's electroweak
theory, which is a cornerstone of the Standard Model. Physicists now
anticipate a Grand Unified Theory (GUT) which can unify the strong force
with the electroweak theory, and a ``Theory of Everything'' (TOE) which will
unite these with gravity.

In the simplest GUT first proposed by Georgi and Glashow\cite{GeorgiGlashow}
(following some work of Pati and Salam\cite{PatiSalam}), the gauge groups
are unified as SU(3)$_{\text{C}}\times $ SU(2)$_{\text{L}}\times $ U(1)$_{%
\text{Y}}\subset $ SU(5) . The running couplings of the three low-energy
gauge groups are predicted to converge at a high scale where the full SU(5)
symmetry is restored. Precision measurements of the gauge couplings at $%
M_{Z} $ by the CERN $e^{+}e^{-}$ collider LEP are consistent with
supersymmetric SU(5)\cite{Amadi,Barger} (because of quantum corrections due
to SUSY particles lying in mass between the weak and GUT scales), but not
with minimal \emph{non}supersymmetric grand unification.%

\miscplot{coupling}{114 318 470 482}%
{Gauge coupling unification.}{%
Gauge coupling unification in (a) the Standard Model and (b) the
MSSM with sparticle threshold at 1 TeV.
Figure reproduced from \cite{Barger}.
}

This fact is suggestive. But whether or not SU(5)\ is the unification group,
and even if nature ``unifies'' without grand unification, as in string
models, SUSY theories generically fix a problem common to nonsupersymmetric
theories. This flaw, the hierarchy problem\cite{Susskind}, comes from
quadratically divergent quantum corrections to masses of scalar particles
such as the SM Higgs (since there is no symmetry in the SM to protect the
scalar masses the way chiral and gauge symmetries, respectively, protect
fermion and vector boson masses). This divergent contribution is due to the
diagrams shown in Fig.~\ref{fig:nonrenorm}. If there were no new physics
whatsoever beyond the SM, then this divergence would be uncontrolled. So the
SM must break down at \emph{some} scale, where new physics intervenes. Very
general considerations suggest, for instance, that field theories in which
gravitation is not incorporated cannot be extrapolated beyond the Planck
scale $\left( \sim 10^{19}\text{GeV}\right) $, where gravitational
interactions are of the same strength as gauge interactions. New physics
must enter here, or perhaps at the somewhat lower gauge coupling unification
scale $\sim 10^{16}$ GeV.

Perturbative unitarity arguments\cite{DicusMather,LeeQuigg} require the
Higgs mass to be less than a few hundred GeV. Keeping the Higgs mass at this
scale while integrating the quadratic divergence up to, say, $\Lambda \sim
10^{16}$ GeV would require introducing a counterterm into the Lagrangian
that must be fine-tuned to an accuracy of (100 GeV/$\Lambda $)$^{2}\sim
10^{-26}$. While not a logical impossibility, this is repugnant to many
physicists. In order to avoid an unnatural fine-tuning, the scale of new
physics cannot be much larger than $\Lambda \sim 1$~TeV. If supersymmetric
particles have masses at this scale, then they can do the job since by the
nonrenormalization theorem of Grisaru, Rocek and Siegel\cite{GrisRoSieg},
the scalar masses in SUSY theories are not quadratically divergent even if
SUSY is spontaneously or softly broken\footnote{%
``Soft'' SUSY breaking terms are just those that do not lead to the
reappearance of quadratic divergences in the scalar masses; these include
explicit mass terms for scalars (e.g., to lift the quark/squark degeneracy),
trilinear scalar interactions (important in the stop sector), and others\cite
{GirGri}.}. The quadratically divergent contributions from scalars in loops
are exactly cancelled by contributions from the scalars' superpartners above
the SUSY restoration scale, illustrated in Fig.~\ref{fig:nonrenorm}.
Supersymmetry is the only symmetry known which offers generic protection to
the scalar masses from the quadratic instability to radiative corrections.%

\miscplot{nonrenorm}{223 566 367 646}%
{Nonrenormalization theorem}{%
Diagrams relevant to the SUSY nonrenormalization theorem.
The scalar loops diverge quadratically, as do the
fermion loops. Under supersymmetry these contributions cancel,
and the result diverges only logarithmically.
}

Of course, supersymmetry is not the only extension to the SM which has been
proposed. Popular alternatives in the literature include technicolor\cite
{Susskind} and compositeness\cite{compositeness}. These are attractive
theories but they seem to be ruled out in their simplest forms. They are
strongly coupled theories, like QCD, and are often not amenable to
calculation\footnote{%
the unreasonable \emph{in}effectiveness of mathmatics in the natural
sciences, to paraphrase Eugene Wigner\cite{wigner}.}. SUSY on the other hand
is a weakly coupled theory whose couplings mainly derive from the SM gauge
couplings in a canonical way. SUSY diagrams are calculable in perturbation
theory, and in fact a wide range of computer tools exists for this purpose 
\cite{isajet2,spythia,susygen,herwig}.

That supersymmetry addresses such a wealth of physics issues with such
economy brings to mind Hertz' comment on Maxwell's equations quoted by
Freeman Dyson\cite{Dyson} (from \cite{Muenster}):

\begin{quote}
One cannot escape the feeling that these mathematical formulae have an
independent existence and an intelligence of their own, that they are wiser
than we are, wiser even than their discoverers, that we get more out of them
than was originally put into them.
\end{quote}

\section{The minimal supersymmetric standard model}

\begin{table}[tb] \centering%
\begin{tabular}{c|c|c|c|}
\cline{2-4}
& {\small spin 0} & {\small spin 1/2}$\mathstrut _{\mathstrut }^{\mathstrut
} $ & {\small spin 1} \\ \hline
\multicolumn{1}{|c|}{{\small 
\begin{tabular}{c}
left-handed \\ 
matter
\end{tabular}
}} & $\left( 
\begin{array}{c}
\tilde{u}_{L} \\ 
\tilde{d}_{L}
\end{array}
\right) ,\left( 
\begin{array}{c}
\tilde{\nu}_{L} \\ 
\tilde{e}_{L}
\end{array}
\right) _{\mathstrut }^{\mathstrut }$ & $\left( 
\begin{array}{c}
u_{L} \\ 
d_{L}
\end{array}
\right) ,\left( 
\begin{array}{c}
\nu _{eL} \\ 
e_{L}
\end{array}
\right) $ &  \\ \cline{1-3}
\multicolumn{1}{|c|}{{\small 
\begin{tabular}{c}
right-handed \\ 
matter
\end{tabular}
}} & $\tilde{u}_{R},\tilde{d}_{R},\tilde{e}_{R}$ & $u_{R},d_{R},e_{R}%
\mathstrut _{\mathstrut }^{\mathstrut }$ &  \\ \cline{1-3}
\multicolumn{1}{|c|}{{\small 
\begin{tabular}{c}
Higgs \\ 
sector
\end{tabular}
}} & $
\begin{array}{c}
\left( 
\begin{array}{c}
H_{d}^{0} \\ 
H_{d}^{-}
\end{array}
\right) ,\left( 
\begin{array}{c}
H_{u}^{+} \\ 
H_{u}^{0}
\end{array}
\right) ^{\mathstrut } \\ 
\quad \quad \Longrightarrow h,A,H^{0},H^{\pm }\mathstrut _{\mathstrut
}^{\mathstrut }
\end{array}
$ & $
\begin{array}{c}
\mathstrut \\ 
\widetilde{H}_{u},\widetilde{H}_{d},\widetilde{B},\widetilde{\mathit{W}} \\ 
\Longrightarrow
\end{array}
$ &  \\ \cline{1-1}\cline{1-2}\cline{4-4}
\multicolumn{1}{|c|}{{\small 
\begin{tabular}{c}
electroweak \\ 
gauge bosons
\end{tabular}
}} &  & $
\begin{array}{c}
\widetilde{Z}_{i}\quad \left( i=1,2,3,4\right) , \\ 
\widetilde{W}_{j}^{\pm }\quad \left( j=1,2\right) \\ 
\mathstrut \\ 
\mathstrut
\end{array}
$ & $
\begin{array}{c}
B,\left( 
\begin{array}{c}
W^{1} \\ 
W^{2} \\ 
W^{3}
\end{array}
\right) _{\mathstrut }^{\mathstrut } \\ 
\quad \Longrightarrow \gamma ,Z,W^{\pm }
\end{array}
$ \\ \cline{1-1}\cline{3-4}
\multicolumn{1}{|c|}{\small gluon} &  & $\tilde{g}\mathstrut _{\mathstrut
}^{\mathstrut }$ & $g$ \\ \hline
\end{tabular}
\caption{MSSM particle content.
Only the first generation of matter particles is shown; the second and third
generations are replicas of this.\label{tab:MSSM}}%
\end{table}%

An immediate consequence of the spinorial nature of the supersymmetry
generator is that SUSY transforms bosons and fermions into each other. A
supersymmetric theory must contain for each fermion a boson with identical
quantum numbers, and vice versa. Spin-$\frac{1}{2}$ matter fermions get
spin-0 partners, and the spin-0 Higgs bosons and spin-1 gauge bosons get
spin-$\frac{1}{2}$ partners. The minimal extension of the SM needed to do
the job has the particle content listed in Table \ref{tab:MSSM}. This
generic SUSY model is called the Minimal Supersymmetric Standard Model
(MSSM). Supersymmetric particles are marked with tildes. By convention,
fermionic superpartners of SM bosons take the suffix ``ino,'' and scalar
counterparts of SM fermions have the prefix ``s''\ (short for ``SUSY''). The
term ``sparticle'' refers to any of these supersymmetric particles.

Matter fermions are represented by Dirac spinors and have four degrees of
freedom, while (complex)\ scalars have only two, so each fermion $f$ is
associated with a pair of scalars. Conventionally these are chosen to be the
sparticles corresponding to the fermion's two helicity states, and are
written as $\tilde{f}_{L}$ and $\tilde{f}_{R}$. (This notation is perhaps
confusing since scalars of course have no chirality; it is still true,
though, that only $\tilde{f}_{L}$ couples to the $W$ (as also only $f_{L}$%
).) The particle studied in this thesis is the lighter supersymmetric
partner of the top quark (a linear combination of $\tilde{t}_{L}$ and $%
\tilde{t}_{R}$), variously called a top squark, scalar top, or stop. For
brevity, we use ``stop'' in the present work.

The only new non-supersymmetric content in Table \ref{tab:MSSM} is an extra
Higgs doublet, placing the MSSM in the class of Two Higgs Doublet Models
(2HDMs). In the Standard Model, three of the Higgs doublet's (real)\ degrees
of freedom go to supply longitudinal polarization for the electroweak gauge
bosons, leaving a single neutral scalar Higgs. In 2HDMs similarly $8-3=5$
states remain for physical particles: two CP even Higgs bosons, a CP odd
Higgs boson, and a pair of charged Higgs.

This extra Higgs doublet is a consequence of consistently applying
supersymmetric field theory. The technique involved---the so-called
superfield formalism---is too lengthy to go into here, but many standard
treatments exist (see \cite{DRTJones} and references therein). One writes
down the most general supersymmetric Lagrangian including soft SUSY breaking
terms (which lift the mass degeneracy of superpartners) consistent with
Poincar\'{e} and gauge invariance. With this treatment it is easy to see
that SUSY requires separate Higgs fields to couple to up-type and down-type
quarks. An immediate consequence of this fact is the introduction of two new
fundamental parameters. One is the ratio of the VEVs for the two Higgs
scalars, $\tan \beta =\upsilon _{\text{u}}/\upsilon _{\text{d}}$. The other
is a higgsino mixing parameter $\mu $ which parameterizes the coupling $\mu 
\widetilde{H}_{u}\widetilde{H}_{d}$; $\mu $ is required to be non-zero to
prevent problems due to a light higgsino.

\newpage
The fermionic partners of the SM photon (photino), $Z$ boson (zino) and two
neutral Higgses (higgsinos)\ all share the same SU(3)$\times $U(1)$_{\text{em%
}}$ quantum numbers and will, therefore, mix if electroweak symmetry is
broken. The resulting states, called neutralinos, are denoted ${\widetilde{Z}%
_{1}}$, ${\widetilde{Z}_{2}}$, ${\widetilde{Z}_{3}}$ and ${\widetilde{Z}_{4}}
$ in order of increasing mass\footnote{%
There are two systems of notation for particles in the electroweak
gaugino/higgsino sector. In this thesis we call the neutral members of this
group ${\widetilde{Z}_{i}}$ ($i=1,2,3,4)$ and the charged members ${%
\widetilde{W}_{j}^{\pm }}$ ($j=1,2)$ (although we usually suppress the
charge sign and just write ${\widetilde{W}_{j})}$ out of prejudice for the
gaugino character of ${\widetilde{Z}_{1,2}}$ and ${\widetilde{W}_{1}}$
expected in the mSUGRA models introduced below. In other works these neutral
and charged ``inos'' are written as $\tilde{\chi}_{i}^{0}$ and $\tilde{\chi}%
_{j}^{\pm }$, respectively.}. For many phenomenological considerations, it
is significant whether the ${\widetilde{Z}_{1}}$ has a predominantly gaugino
or higgsino character. Since we typically set supersymmetric branching
fractions to 100\% for the decays we study here, this distinction is largely
unimportant for our work. As with the neutralinos, the superpartners of the $%
W$ and the charged Higgs also mix to form two chargino states, the lighter ${%
\widetilde{W}_{1}}$ and the heavier ${\widetilde{W}_{2}}$. The fermionic
content that the MSSM adds to the SM (gauginos+higgsinos)\ is free of chiral
anomalies.

General techniques for the construction of supersymmetric particle
Lagrangians are discussed in \cite{tataBrazil}. In their most general form,
these SUSY Lagrangians admit gauge-invariant renormalizable lepton- and
baryon-number violating interactions, in contrast to the Standard Model. The
completely unconstrained MSSM scheme is minimal with respect to particle
content, but maximal in terms of parameters---106 new parameters are
required in addition to the SM's 18\cite{DimoSutter}. These new parameters
are strongly constrained by current data on CP violation, flavor-changing
neutral currents, and lepton flavor non-conservation. Setting these possible
phases are all to zero, the resulting theory has 30 new parameters. The MSSM
referred to in this thesis is this 30+18 parameter model.

$B$ and $L$ conservation leads to a further discrete symmetry of the
renormalizable terms in the Lagrangian, $R$-parity invariance, with quantum
number $R=\left( -1\right) ^{3B+L+2S}$ (where $S=$ spin). All SM particles
(plus the second Higgs doublet) are $R$-even, and all superpartners are $R$%
-odd. If $R$-parity is a good symmetry then sparticles must be pair-produced
by collision of SM particles, and a decaying sparticle must have an odd
number of sparticles among its decay products. Therefore there must be a
lightest supersymmetric particle (LSP) which is stable. Limits on long-lived
colored or charged particles require the LSP to be weakly interacting and
electrically neutral\cite{LSPneutral}. The ${\widetilde{Z}_{1}}$ is
generally taken to be the LSP. (The only other LSP candidate is the scalar
neutrino $\tilde{\nu}$, but the sneutrino is disfavored if the LSP is also
to constitute the cold dark matter presumed to lie in the galactic halo\cite
{cosmoLSP}.) In SUSY events at particle colliders, this LSP will escape the
detector without interacting, leading to the classic SUSY signature of
missing transverse energy $\NEG{E}\mathstrut _{\top }^{\text{ }}$.

If SUSY is to solve the hierarchy problem, then the masses of the
superpartners of the MSSM cannot be much greater than 1 TeV, and so particle
colliders currently in operation (especially the Fermilab Tevatron) or under
construction (the CERN\ LHC) have a good chance of finding some evidence for
their existence. This accessible scale of new particle states is small
compared to the scale imagined for the mechanisms underlying supersymmetry
(from 10$^{10}$ to 10$^{19}$ GeV), and these theories are generically called
``low-energy supersymmetry.''

\section{Constrained SUSY models}

The MSSM is completely agnostic about physics above the TeV scale of new
particle states. The many parameters mentioned in the previous section are
so-called ``soft'' SUSY breaking parameters, and are written down from very
general considerations of the supersymmetric Lagrangian. The parameter space
of the general MSSM is too large for viable phenomenological studies, and
one therefore turns to more restricted models. Further constraints among
these parameters require a theory of physics at energies beyond 1 TeV. There
being a paucity of data on physics at such scales, no complete and
compelling supersymmetric theory has emerged as yet.

Constraints can be found by considering the nature of supersymmetry
breaking. SUSY must of course be broken since no sparticles degenerate in
mass with their ordinary partners (e.g., scalar electrons) have been
observed. SUSY breaking relying only on the MSSM content of the theory (or
even on just the TeV scale with more particles) seems to be infeasible, so
models invoke a ``hidden'' sector of particles which have no SM gauge
interactions.\ In supergravity models, SUSY is broken in this hidden sector,
which communicates with the visible sector through gravitational
interactions alone. One obtains a simple structure which one assumes is
valid at a high energy scale ($M_{\text{GUT}}$ or $M_{\text{planck}}$) and
uses renormalization group equations (RGE) to calculate low energy MSSM
parameters. The high energy parameters are usually universal (for instance
with gravity-mediated SUSY breaking), and a GUT ansatz affords further
simplification by relating gaugino masses.

The minimal supergravity (mSUGRA) models widely used in phenomenological
studies are prepared according to this prescription\cite{mSUGRA}. mSUGRA
depends on only four continuous parameters and a sign in addition to the 18
parameters of the SM. These are generally taken to be a universal scalar
mass $m_{0}$, a universal gaugino mass $m_{1/2}$, a universal trilinear
interaction $A_{0}$ (sfermion-Higgs-sfermion), the Higgs VEV ratio $\tan
\beta $, and the sign of the Higgsino mixing parameter $\mu $: 
\begin{eqnarray}
m_{0}\text{, }m_{1/2}\text{, }A_{0}\text{, }\tan \beta \text{, sign}\left(
\mu \right) 
\label{eq:sugra:params}%
\end{eqnarray}
all input at the GUT scale except for $\tan \beta $, which is a weak-scale
quantity. The Higgs bosons, which are part of chiral supermultiplets, share
the common scalar mass $m_{0}$ at the high scale, and are renormalized due
to gauge interactions like the doublet sleptons. The Higgs doublet that
couples to the top family gets very large negative contributions to its
squared mass $m_{H_{u}}^{2}$ due to the large top Yukawa coupling (discussed
below) which can easily be driven to negative values leading automatically
and naturally to radiative electroweak symmetry breaking (EWSB).\footnote{%
The electroweak symmetry breaking is automatic and natural, but not
miraculous; one still has to input the weak scale by hand into the high
energy parameters of the theory.} In mSUGRA the magnitude $\left| \mu
\right| $ is fixed by requiring EWSB to reproduce the measured $Z$ boson
mass.

mSUGRA has many distinctive features, and its phenomenology has been studied
extensively. We collect a few facts here which will be useful in our work.
Because the high-scale mass parameters are universal and the RGE depend
mainly on the known gauge couplings, one gets generic mass relationships
such as 
\begin{eqnarray}
m_{\tilde{q}}^{2}=m_{0}^{2}+m_{q}^{2}+(5-6)m_{1/2}^{2}+D\text{-terms} 
\end{eqnarray}
where the D-term scale is $M_{Z}^{2}/2$. The squark, slepton and gluino
masses obey 
\begin{eqnarray}
m_{\tilde{q}}^{2}=m_{\tilde{\ell}}^{2}+\left( 0.7-0.8\right) m_{\tilde{g}%
}^{2}%
\label{eq:sugra:masses}%
\end{eqnarray}
(ignoring quark and lepton masses). The lighter neutralinos and lighter
chargino are mostly gaugino (rather than higgsino) and one generally has 
\begin{eqnarray}
2m_{{\widetilde{Z}_{1}}}\approx m_{{\widetilde{W}_{1}}}\approx m_{\tilde{g}%
}/3.%
\label{eq:sugra:winozino}%
\end{eqnarray}
The relations (\ref{eq:sugra:winozino}) hold because gaugino masses are
assumed to unify at the GUT scale.

mSUGRA is just one of a large and growing set of models intended to capture
physics at the high scale. There are theories that are more constrained,
such as certain superstring-inspired models, in which only one or two free
parameters appear. There has also been recent interest in models which vary
the different mSUGRA assumptions on universality at the GUT scale. For
instance, this may happen if masses unify at $M_{\text{Planck}}$ and become
non-universal as they run down to $M_{\text{GUT}}$, or if the unifying group
is larger than SU(5)\ and admits additional mass contributions. String
models, hypercolor models, and models with novel SUSY breaking mechanisms
can also exhibit non-universality. Reference \cite{BDQT} surveys the gross
phenomenology of these models. Because the light stop sector depends on
relatively few MSSM parameters, our results are not particularly sensitive
to the actual physics at the high scale. Our results, as we will see, will
broadly depend only on how the stop is assumed to decay.

\chapter{The Top Squark%
\label{chap:stop}%
}

\section{The light stop}

The top quark has recently been detected and its mass measured in the lepton
+ jets\cite{CDFtoplj,D0toplj1,D0toplj2}, dilepton\cite{CDFtopll,D0topll} and
all-jets\cite{CDFtopjets} channels. A concise summary of these results is
presented in \cite{topMass} which derives a combined value of $%
m_{t}=174.3\pm 3.2\left( \text{statistical}\right) \pm 4.0\left( \text{%
systematic}\right) $ GeV. For the purpose of our calculations, we adopt $%
m_{t}=175$ GeV.

With such an extraordinary mass, the top family Yukawa interactions are
large and comparable to the electroweak gauge interactions.\ In
renormalizing squark masses from the unification scale down to the physical
mass scale, this top Yukawa coupling depresses the diagonal masses of the
left and right stop states ${\tilde{t}_{L}}$ and ${\tilde{t}_{R}}$ relative
to the masses of the generation 1 and 2 squarks and also allow substantial
mixing between the two states\cite{EllisRudaz,IKKT,IbanLop}. The stop masses
come from Higgs superpotential terms 
\begin{eqnarray}
f_{\text{MSSM}}\ni \mu \widehat{h}_{u}^{0}\widehat{h}_{d}^{0}+f_{t}\widehat{t%
}\widehat{h}_{u}^{0}\widehat{T}^{C}%
\label{eq:HiggsSP}%
\end{eqnarray}
($\widehat{h}_{u}^{0}$ and $\widehat{h}_{d}^{0}$ are the up- and down-type
Higgs superfields, $\widehat{t}$ and $\widehat{T}^{C}$ are left and right
top superfields) and soft SUSY breaking terms 
\begin{eqnarray}
\mathcal{L}_{\text{soft SUSY breaking}}\ni A_{t}f_{t}\tilde{t}_{L}H_{u}^{0}%
\tilde{t}_{R}^{\dagger }-m_{\tilde{t}_{L}}^{2}\tilde{t}_{L}^{\dagger }\tilde{%
t}_{L}-m_{\tilde{t}_{R}}^{2}\tilde{t}_{R}^{\dagger }\tilde{t}_{R}.%
\label{eq:trilinear}%
\end{eqnarray}

The stop masses are (assuming universality as in mSUGRA) parameterized as 
\begin{eqnarray}
\begin{array}{ccl}
m_{{\tilde{t}_{L}}}^{2} & =_{\mathstrut _{\mathstrut _{\mathstrut
}}}^{\mathstrut ^{\mathstrut ^{\mathstrut }}} & \stackrel{\text{{\small %
GUT-scale\ mass}}}{\overbrace{m_{0}^{2}}}+\stackrel{\text{{\small gauge RGE}}%
}{\overbrace{Fm_{1/2}^{2}}}-\stackrel{\text{{\small Yukawa RGE}}}{\overbrace{%
2\left( f_{t}^{2}\Delta _{t}+f_{b}^{2}\Delta _{b}\right) m_{0}^{2}}}+%
\stackrel{\text{{\small EWSB\ mass}}}{\overbrace{m_{t}^{2}}} \\ 
& \mathstrut _{\mathstrut _{\mathstrut _{\mathstrut }}}^{\mathstrut
^{\mathstrut ^{\mathstrut }}} & \qquad \qquad \qquad \qquad \qquad -%
\stackrel{\text{\textit{D} term}}{\overbrace{\left( -\cos 2\beta \right)
\left( \frac{1}{2}-\frac{2}{3}\sin ^{2}\theta _{W}\right) M_{Z}^{2}}} \\ 
& =_{\mathstrut _{\mathstrut _{\mathstrut }}}^{\mathstrut ^{\mathstrut
^{\mathstrut }}} & m_{\tilde{u}_{L}}^{2}+m_{t}^{2}-2\left( f_{t}^{2}\Delta
_{t}+f_{b}^{2}\Delta _{b}\right) m_{0}^{2}, \\ 
m_{{\tilde{t}_{R}}}^{2} & =_{\mathstrut _{\mathstrut _{\mathstrut
}}}^{\mathstrut ^{\mathstrut ^{\mathstrut }}} & m_{\tilde{u}%
_{R}}^{2}+m_{t}^{2}-4f_{t}^{2}\Delta _{t}m_{0}^{2}.
\end{array}
\end{eqnarray}
where the 1$^{\text{st}}$ and 2$^{\text{nd}}$ generation\ left- and
right-handed squark masses at the weak scale are 
\begin{eqnarray}
\begin{array}{ccl}
m_{\tilde{u}_{L}}^{2} & =_{\mathstrut _{\mathstrut _{\mathstrut
}}}^{\mathstrut ^{\mathstrut ^{\mathstrut }}} & m_{0}^{2}+Fm_{1/2}^{2}-%
\left( -\cos 2\beta \right) \left( \frac{1}{2}-\frac{2}{3}\sin ^{2}\theta
_{W}\right) M_{Z}^{2}, \\ 
m_{\tilde{u}_{R}}^{2} & =_{\mathstrut _{\mathstrut _{\mathstrut
}}}^{\mathstrut ^{\mathstrut ^{\mathstrut }}} & m_{0}^{2}+Fm_{1/2}^{2}-%
\left( -\cos 2\beta \right) \left( \frac{2}{3}\sin ^{2}\theta _{W}\right)
M_{Z}^{2}.
\end{array}
\end{eqnarray}
At the GUT scale the stops get contributions from the universal scalar mass $%
m_{0}$. Gauge interactions in the RGE (predominantly QCD) add the term
proportional to the gaugino mass, where the coefficient $F$ has a typical
value of $5-6$. The top and bottom Yukawa couplings are 
\begin{eqnarray}
\begin{array}{ccl}
f_{t}^{2} & =_{\mathstrut _{\mathstrut _{\mathstrut }}}^{\mathstrut
^{\mathstrut ^{\mathstrut }}} & \left. g^{2}m_{t}^{2}\right/ \left(
2M_{W}^{2}\sin ^{2}\beta \right) \approx 1.01/\sin ^{2}\beta , \\ 
f_{b}^{2} & =_{\mathstrut _{\mathstrut _{\mathstrut }}}^{\mathstrut
^{\mathstrut ^{\mathstrut }}} & \left. g^{2}m_{b}^{2}\right/ \left(
2M_{W}^{2}\cos ^{2}\beta \right) \approx 0.00082/\cos ^{2}\beta 
\end{array}
\end{eqnarray}
with $g$ the weak coupling. The quantities $\Delta _{t,b}$ are calculated
from the RGE and take values on the order of $0.1$. For small $\tan \beta $
the top Yukawa $f_{t}$ dominates and $m_{{\tilde{t}_{R}}}<m_{{\tilde{t}_{L}}}
$ since ${\tilde{t}_{R}}$ gets twice as much depression as ${\tilde{t}_{L}}$%
; these two masses approach each other as $\tan \beta \rightarrow m_{t}/m_{b}
$. The usual EWSB mechanism adds the top-flavor mass term $m_{t}^{2}$. The
factor $\cos 2\beta =\left( 1-\tan ^{2}\beta \right) \left/ \left( 1+\tan
^{2}\beta \right) \right. $ in the $D$-term is negative for $\tan \beta >1$,
so the $D$-terms depress the diagonal masses.

The Yukawa interaction also mixes ${\tilde{t}_{L}}$ and ${\tilde{t}_{R}}$
and the stop mass matrix from interactions (\ref{eq:HiggsSP}) and (\ref
{eq:trilinear}) is\cite{StopMatrix} 
\begin{eqnarray}
\mathcal{M}^{2}=\left[ 
\begin{array}{cc}
m_{{\tilde{t}_{L}}}^{2} & -a_{t}m_{t} \\ 
-a_{t}m_{t} & m_{{\tilde{t}_{R}}}^{2}
\end{array}
\right] 
\end{eqnarray}
with $a_{t}=A_{t}-\mu \cot \beta $. Diagonalizing this gives mass
eigenvalues 
\begin{eqnarray}
m_{{\tilde{t}_{1,2}}}^{2} &=&\frac{m_{{\tilde{t}_{L}}}^{2}+m_{{\tilde{t}_{R}}%
}^{2}}{2}\mp \sqrt{\left( \frac{m_{{\tilde{t}_{L}}}^{2}-m_{{\tilde{t}_{R}}%
}^{2}}{2}\right) ^{2}+a_{t}^{2}m_{t}^{2}} \\
&=&m_{\tilde{q}}^{2}+m_{t}^{2}-\left( 3f_{t}^{2}\Delta _{t}+f_{b}^{2}\Delta
_{b}\right) m_{0}^{2}%
\nonumber%
\\
&&\mp \sqrt{\left( f_{t}^{2}\Delta _{t}m_{0}^{2}-\left( 1-\frac{8}{3}\sin
^{2}\theta _{W}\right) \frac{-\cos 2\beta }{4}M_{Z}^{2}\right) ^{2}+\left(
A_{t}-\mu \cot \beta \right) ^{2}m_{t}^{2}}%
\nonumber%
\end{eqnarray}
and the physically propagating eigenstates 
\begin{eqnarray}
\left( 
\begin{array}{c}
{\tilde{t}_{1}} \\ 
{\tilde{t}_{2}}
\end{array}
\right) =\left[ 
\begin{array}{cc}
\cos \theta _{t} & -\sin \theta _{t} \\ 
\sin \theta _{t} & \cos \theta _{t}
\end{array}
\right] \left( 
\begin{array}{c}
{\tilde{t}_{L}} \\ 
{\tilde{t}_{R}}
\end{array}
\right) 
\end{eqnarray}
rotated by the stop mixing angle 
\begin{eqnarray}
\tan \theta _{t}=\frac{m_{{\tilde{t}_{L}}}^{2}-m_{{\tilde{t}_{1}}}^{2}}{%
m_{t}\left( A_{t}-\mu \cot \beta \right) }. 
\end{eqnarray}
In the absence of mixing, $\theta _{t}\rightarrow \pi /2$ and ${\tilde{t}%
_{1}=\tilde{t}_{R}}$ in models with a universal mass at the high scale.

For moderate $\tan \beta $ the stop mass splitting is dominated by the
off-diagonal terms in the mass matrix.\ The unification scale trilinear
coupling parameter $A_{0}$ can be freely adjusted, and light stops can
usually be arranged irrespective of the value of the other universal mass
parameters $m_{0}$ and $m_{1/2}$; such parameter sets can generally be made
consistent with experimental constraints (particularly those on $M_{Z}$ and $%
m_{H}$)\cite{Regina}.\ In fact, ${\tilde{t}_{1}}$ masses can be driven low
enough that model builders have to be careful to insure that the stop does
not become the LSP or, worse, tachyonic leading to dangerous color-breaking
minima.

\subsection{Theoretical prejudices%
\label{sec:stop:theo}%
}

We have seen that it is possible for the top squark to be light. There are
also arguments that directly favor a light ${\tilde{t}_{1}}$.\ Chief among
these is electroweak baryogenesis, a mechanism for generating the observed
baryon number of the universe at the electroweak phase transition\cite
{EWbaryo}. The original baryogenesis idea of Sakharov\cite{Sakharov}
involving baryon number violation, C and CP violation and thermal
nonequilibrium seems not to work in the Standard Model. The CP-violating
phase in the Kobayashi-Maskawa matrix is too small to allow enough baryon
number production in the symmetric phase, and experiment rules out a Higgs
scalar light enough to generate a strongly first order phase transition
which would lock in the generated baryons. SUSY, though, has many extra
CP-violating phases. And a light (scalar) stop strongly coupled to the Higgs
through its large Yukawa coupling can provide for the first order phase
transition while allowing the Higgs to be heavier. Detailed calculations\cite
{CarQuirWag,Delepine} indicate that the stop mass must lie in the range $100$
GeV $\lesssim m_{{\tilde{t}_{1}}}\lesssim 160$ GeV for this mode of
electroweak baryogenesis to succeed. We will find that this range is quite
accessible to the Fermilab Tevatron using the searches outlined in this
thesis.

Let us collect here a few other theoretical considerations related to the
stop. First, as mentioned above, the LSP ${\widetilde{Z}_{1}}$ is a favored
candidate for the dark matter determined to be present in the halo of our
galaxy. In order for these relic neutralinos to have an appropriate density,
it is necessary that they cannot have annihilated via ${\widetilde{Z}_{1}+%
\widetilde{Z}_{1}\rightarrow }$ SM particles at too great a rate. If ${%
\tilde{t}_{1}}$ is not much heavier than ${\widetilde{Z}_{1}}$, then it is
also possible to spoil the relic density through co-annihilation\cite
{cosmoLSPbjd}. The authors of \cite{cosmoLSPbjd} have studied this process,
and concluded that maintaining the favored relic density $0.1<\Omega _{{%
\widetilde{Z}_{1}}}<0.2$ requires a mass gap $m_{{\tilde{t}_{1}}}-m_{{%
\widetilde{Z}_{1}}}>11$ to 33 GeV. Excluding the small mass gap is good news
for collider experiments, where small $m_{{\tilde{t}_{1}}}-m_{{\widetilde{Z}%
_{1}}}$ means poor detection efficiency.

There are also constraints on the stop sector coming from precision
electroweak measurements. One of these is the $\rho $ parameter\cite
{DreesHagiwara}, which is a measure of the ratio of charged current to
neutral current interaction strength. The close agreement between LEP\ II
experimental determinations and the Standard Model $\Delta \rho =0$ means
that $\tilde{t}_{L}$, which couples to the $W$, must be quite heavy. A
recent study\cite{rhoParam} finds that in mSUGRA one needs $m_{\tilde{t}%
_{L}}>275$ (310) GeV for $\mu >0$ ($\mu <0$). $\rho $ parameter limits on
the ${\widetilde{W}_{1}}$ and ${\widetilde{Z}_{1}}$ masses are weaker than
the current direct experimental limits.

\miscplot{penguins}{136 614 488 686}%
{Diagrams for $b\to s\gamma$}{%
contributions to the decay $b\to s\gamma$ from the Standard Model (SM),
two Higgs double models (2HDM), and supersymmetric models (SUSY).}

In 2HDM models with a light charged Higgs the rare decay $b\rightarrow
s\gamma $ may require a light stop as well. This decay has been measured\cite
{bsgCleo} to have a rate close to its SM predicted value. A light $H^{-}$
enhances the decay by allowing a charged Higgs diagram analogous to the SM $%
W^{-}$ diagram, as in Fig.~\ref{fig:penguins}, while a SUSY diagram with a
chargino and light stop in the loop decreases the rate and can bring it back
into line with the SM rate\cite{bsgSUSY,Okada}. This is generally not a
problem in models, as the charged Higgs is typically heavy. However, the $%
\rho $ parameter data allow an mSUGRA $H^{-}$ as light as 140 GeV ($\mu <0$) 
\cite{rhoParam}.

\section{Stop production and decay%
\label{sec:stop:prod}%
}

Stops are color triplets, as are top quarks, and at a hadron collider such
as the Fermilab Tevatron will be strongly produced in ${\tilde{t}_{1}^{\ast }%
\tilde{t}_{1}}$ pairs (to conserve R-parity) through gluon fusion and $q\bar{%
q}$ annihilation. The tree level production cross-section $\sigma _{{\tilde{t%
}_{1}^{\ast }\tilde{t}_{1}}}$ depends only on the mass $m_{{\tilde{t}_{1}}}$%
. Figure~\ref{fig:stopprod} shows this tree level $\sigma _{{\tilde{t}%
_{1}^{\ast }\tilde{t}_{1}}}$ as a function of $m_{{\tilde{t}_{1}}}$ for the
Fermilab Tevatron $p\bar{p}$ collider at a center of mass energy $\sqrt{s}%
=2.0$ TeV. These cross-sections were produced by the program {\small ISAJET}
(see below) using the \texttt{CTEQ2L\footnote{%
We have checked that our results are not affected by using more modern
parton distribution functions than the \texttt{CTEQ2L} of our original
analysis; this is as expected since the relatively heavy stop pairs are
produced at high $x$.}} parton distribution functions\cite{CTEQ2L}.\ For
comparison, the top quark pair production cross section for $2.0$ TeV $p\bar{%
p}\rightarrow t\bar{t}$ is $\sigma _{t\bar{t}}=6100$ fb when calculated by
the same means.

\begin{table}[b] \centering%
$
\begin{tabular}{|c|c|}
\hline
$m_{{\tilde{t}_{1}}}$ (GeV) & $K$ \\ \hline
70 & 1.41 \\ 
110 & 1.30 \\ 
150 & 1.19 \\ 
190 & 1.11 \\ \hline
\end{tabular}
$\caption{K factors for Tevatron stop pair production.\label{tab:Kfactors}}%
\end{table}%

The Next to Leading Order (NLO) corrections to the tree-level $\sigma _{{%
\tilde{t}_{1}^{\ast }\tilde{t}_{1}}}$ have been calculated\cite{beenakker}.
The authors of \cite{beenakker} used mSUGRA-inspired parameters to calculate
higher order corrections for the Tevatron $p\bar{p}$ collider at $\sqrt{s}%
=1.8$ TeV. The corrections are positive and substantial. Table \ref
{tab:Kfactors} lists the K factors they obtained. We use the lowest order
cross-sections for both signal and background in the present work.

\begin{table}[b] \centering%
$
\begin{tabular}{|c|c|c|}
\hline
particle & mass (GeV) & notes \\ \hline
$\tilde{q}$ & 220 & \cite{D0squark,CDFsquark} \\ 
$\tilde{g}$ & 173 & \cite{D0gluino,CDFsquark} \\ 
${\tilde{t}_{1}}$ & 90 & \cite{OPALstop,CDFscz,D0metjet} \\ 
${\widetilde{W}_{1}}$ & 93 & large $m_{0},$\cite{L3ino,OPALino,ALEPHino} \\ 
${\widetilde{Z}_{1}}$ & 32 & large $m_{0},$\cite{L3ino,OPALino} \\ 
$\tilde{e}$ & 92 & \cite{ALEPH200} \\ 
$H$ & 108 & \cite{Sopczak} \\ \hline
\end{tabular}
$%
\caption{Current mass bounds for particles relevant to this
study.\label{tab:masses}}%
\end{table}%

\miscplot{stopprod}{111 271 486 493}%
{Stop pair production at the Tevatron}{%
Stop pair production cross-section at the Tevatron.}
Current experimental mass limits for SUSY particles of interest here are
listed in Table \ref{tab:masses}. These are generalized lower bounds which
we will use for reference in this thesis. Many of these limits are
correlated, and in some cases higher masses have been excluded for
particular combinations of other SUSY parameters. It is also possible to
find corners of parameter space where these bounds may be evaded.\ See the
indicated references for a fuller discussion. The ${\tilde{t}_{1}}$, ${%
\widetilde{W}_{1}}$ and ${\widetilde{Z}_{1}}$ experimental determinations
are discussed briefly in Section 1.2.3.

Given these masses, the possible light ${\tilde{t}_{1}}$ decays in SUSY
models without exotic particle content are \cite{BDGGT} the 2-body
tree-level modes ${\tilde{t}_{1}}\rightarrow t{\widetilde{Z}_{1}}$ and ${%
\tilde{t}_{1}}\rightarrow b{\widetilde{W}_{1}}$, the 2-body loop mode ${%
\tilde{t}_{1}}\rightarrow c{\widetilde{Z}_{1}}$, the 3-body tree-level modes 
${\tilde{t}_{1}}\rightarrow bW{\widetilde{Z}_{1}}$, ${\tilde{t}_{1}}%
\rightarrow b{\tilde{\ell}}\nu $ and ${\tilde{t}_{1}}\rightarrow b\ell {%
\tilde{\nu}}$, and the four-body decay ${\tilde{t}_{1}}\rightarrow
bff^{\prime }{\widetilde{Z}_{1}}$. Figure \ref{fig:decays1} diagrams the
decays analyzed at length in this thesis, and Fig.~\ref{fig:dknot} displays
the others. Table \ref{tab:decays} presents a summary of these modes.

\miscplot{decays1}{130 620 330 684}%
{Stop decays treated in this thesis.}{%
Decays of the light $\tilde{t}_1$ studied in this thesis.
}

\looseness=2
We could write decays analogous to those of the preceding paragraph with a
gluino $\tilde{g}$ replacing the LSP ${\widetilde{Z}_{1}}$, such as ${\tilde{%
t}_{1}}\rightarrow c\tilde{g}$. The current gluino mass bound (see Table \ref
{tab:masses}) would require a rather heavy stop to open this mode and
overcome phase-space suppression enough to compete with the
(loop-suppressed) ${\tilde{t}_{1}}\rightarrow c{\widetilde{Z}_{1}}$,
especially under the gaugino unification relationship $m_{\tilde{g}}\sim 6m_{%
{\widetilde{Z}_{1}}}$. Even if the reaction did have an appreciable
branching fraction, it would be attended by the cascade $\tilde{g}%
\rightarrow q\bar{q}{\widetilde{Z}_{1}}$ and so would be similar the decay ${%
\tilde{t}_{1}}\rightarrow c{\widetilde{Z}_{1}}$ (which we study at length)
but with more jets and a softer LSP. The other possible gluino modes are
even more suppressed--- ${\tilde{t}_{1}}\rightarrow bW\tilde{g}$ has too
heavy a final state and ${\tilde{t}_{1}}\rightarrow bff^{\prime }\tilde{g}$
has too little phase space. We do not consider gluino decays further.%
\footnote{%
Models with light gluinos\cite{BaerGluino,Clavelli} which could have evaded
detection have been proposed in the literature. \ We do not consider such
models here.}

\miscplot{dknot}{115 311 429 469}%
{Stop decays not treated in this thesis.}{%
Decays of the $\tilde{t}_1$ not studied in this thesis.
}

When $m_{{\tilde{t}_{1}}}>m_{t}+m_{{\widetilde{Z}_{1}}}$ the direct 2-body
decay ${\tilde{t}_{1}}\rightarrow t{\widetilde{Z}_{1}}$ is open, but it will
be strongly suppressed by phase space unless the mass gap is large. This
requires $m_{{\tilde{t}_{1}}}\gg 210$ GeV. From Fig.~\ref{fig:stopprod} the
production cross-section for such stops is less than 150 fb, and at 1/40th
the $t\bar{t}$ production cross-section the likelihood of observing this
mode is remote. We do not consider it further. We also mention in passing
that the supersymmetric top decay $t\rightarrow {\tilde{t}_{1}\widetilde{Z}%
_{1}}$ can open up\cite{Sender} if the stop and neutralino are light enough.
See section 1.2.3 for this.

If $m_{{\tilde{t}_{1}}}>m_{b}+m_{{\widetilde{W}_{1}}}$, the 2-body
tree-level decay ${\tilde{t}_{1}}\rightarrow b{\widetilde{W}_{1}}$ will
dominate. The chargino from ${\tilde{t}_{1}}\rightarrow b{\widetilde{W}_{1}}$
rapidly decays to a neutralino and fermion pair, ${\widetilde{W}_{1}}%
\rightarrow f^{\prime }\bar{f}{\widetilde{Z}_{1}}$, so that ${\tilde{t}_{1}}$
pair production results in ${\tilde{t}_{1}}{}^{\!\ast }{\tilde{t}_{1}}%
\rightarrow b\bar{b}+4\mathrm{\ fermions}+{\widetilde{Z}_{1}}{\widetilde{Z}%
_{1}}$. (We do not consider models with a slepton light enough for ${%
\widetilde{W}_{1}}\rightarrow \tilde{\ell}\nu $, as explained below.) This
top squark event topology is similar to the Standard Model $t\bar{t}$\
pattern $t\bar{t}\rightarrow b\bar{b}+4\mathrm{\ fermions}$, the main
difference being the presence of the ${\widetilde{Z}_{1}}{\widetilde{Z}_{1}}$
pair. We will find that for the range of ${\tilde{t}_{1}}$ masses accessible
with the luminosity upgrades at the Tevatron a real chargino cannot decay
into a real $W$ boson, which provides for another difference from $t\bar{t}$%
\ events. (The chargino will decay to a real $W$ only when $m_{{\widetilde{W}%
_{1}}}-m_{{\widetilde{Z}_{1}}}>m_{W}$. Given the measured $W$ mass and the
experimental limit on the neutralino mass, this requires $m_{{\widetilde{W}%
_{1}}}>115$ GeV, rising to $m_{{\widetilde{W}_{1}}}>160$ GeV as $m_{{%
\widetilde{W}_{1}}}\rightarrow 2m_{{\widetilde{Z}_{1}}}$ (see eqn. \ref
{eq:sugra:winozino}) and we use the $m_{{\widetilde{W}_{1}}}$ limit. As we
will see in Chapter \ref{chap:sbw}, such heavy charginos are not accessible
at the Tevatron in ${\tilde{t}_{1}}\rightarrow b{\widetilde{W}_{1}}$ decays.)

Phenomenologically, ${\tilde{t}_{1}}{}^{\!\ast }{\tilde{t}_{1}}$ events for
the ${\tilde{t}_{1}}\rightarrow b{\widetilde{W}_{1}}$ mode are categorized
according to the leptonic or hadronic nature of the chargino decays just as $%
t\bar{t}$\ events are categorized according to the nature of their $W$
decays. There are (1) dilepton events ${\tilde{t}_{1}}{}^{\!\ast }{\tilde{t}%
_{1}}\rightarrow b\bar{b}\ell ^{+}{\ell ^{\prime }}^{-}\nu \bar{\nu}{%
\widetilde{Z}_{1}}{\widetilde{Z}_{1}}$ whose signature is ($b$)-jets +
dilepton + $\NEG{E}\mathstrut _{\top }^{\text{ }}$, (2) 1-lepton events ${%
\tilde{t}_{1}}{}^{\!\ast }{\tilde{t}_{1}}\rightarrow b\bar{b}q^{\prime }\bar{%
q}\ell \nu {\widetilde{Z}_{1}}{\widetilde{Z}_{1}}$ with signature ($b$)-jets
+ lepton + $\NEG{E}\mathstrut _{\top }^{\text{ }}$, and (3) all-jet events ${%
\tilde{t}_{1}}{}^{\!\ast }{\tilde{t}_{1}}\rightarrow b\bar{b}q^{\prime }\bar{%
q}q^{\prime \prime }\bar{q}^{\prime \prime \prime }{\widetilde{Z}_{1}}{%
\widetilde{Z}_{1}}$ with ($b$)-jets + $\NEG{E}\mathstrut _{\top }^{\text{ }}$%
. The all-jets channel suffers from large QCD backgrounds; the search is
difficult in this channel and we do not consider it further in this study.
The leptonic channels show the same topologies as the corresponding $t\bar{t}
$\ channels, and thus the important backgrounds for these channels are those
usually identified in $t\bar{t}$\ studies, along with $t\bar{t}$\ itself.
These channels are considered in detail in Chapter~\ref{chap:sbw}.

\miscplot{winodecay}{127 566 404 685}%
{Chargino decay diagrams.}{%
Tree-level decays of the chargino. The diagram at left is the
$W^*$-mediated decay, which will lead to $W$-like fermionic branching fractions
if it dominates. The sfermion diagram on the right will enhance the branching
fraction to $\overline{f}'f$ when $\tilde{f}$ is light.}
When stop decays via the chargino ${\tilde{t}_{1}}\rightarrow b{\widetilde{W}%
_{1}}$ then one also has to consider the cascade decay ${\widetilde{W}%
_{1}\rightarrow }f\bar{f}^{\prime }{\widetilde{Z}_{1}}$. We are particularly
interested in the branching fraction to leptons, $\mathcal{B}\left( {%
\widetilde{W}_{1}\rightarrow }\text{ leptons}+{\widetilde{Z}_{1}}\right) $,
since our detection schemes look for these leptons. When the decay proceeds
predominantly via a virtual $W$, as in Fig.~\ref{fig:winodecay}a, then the
chargino branching fractions will be those of the mediating $W$, $\mathcal{B}%
({\widetilde{W}_{1}}\rightarrow e)=\mathcal{B}({\widetilde{W}_{1}}%
\rightarrow \mu )=\mathcal{B}({\widetilde{W}_{1}}\rightarrow \tau )=\mathcal{%
B}(W\rightarrow e)\approx 1/9$. But if some sfermion $\tilde{f}$ is light
enough for the diagram in Fig.~\ref{fig:winodecay}b to compete, then the
branching fraction to the corresponding fermion and its partner ${\widetilde{%
W}_{1}\rightarrow }f\bar{f}^{\prime }{\widetilde{Z}_{1}}$ can be enhanced.
If sleptons are lighter than squarks (but still heavier than ${\widetilde{W}%
_{1})}$ then we can get the branching fraction $\mathcal{B}\left( {%
\widetilde{W}_{1}\rightarrow }\text{ leptons}+{\widetilde{Z}_{1}}\right) \gg
1/9.$ In mSUGRA (and in many other models) we can easily have $m_{\tilde{q}%
}\gg m_{\tilde{\ell}}$ from eqn.~(\ref{eq:sugra:masses}), and one typically
finds an enhancement. This almost always works in our favor, as discussed
below in Chapter \ref{chap:sbw}. The $\tilde{f}^{\ast }$-mediated decay can
also be important when the $W{\widetilde{W}_{1}\widetilde{Z}_{1}}$ coupling
is dynamically suppressed.

In case $m_{{\tilde{t}_{1}}}<m_{b}+m_{{\widetilde{W}_{1}}}$, {the }stop must
go via the 2-body loop decay or the 3- or 4-body decays. In mSUGRA $m_{%
\tilde{\ell}_{L}}>m_{\tilde{\ell}_{R}}$ and experimentally $m_{\tilde{\ell}%
_{R}}>80$ GeV so we will not consider ${\tilde{t}_{1}}\rightarrow b{\tilde{%
\ell}}\nu $ or ${\tilde{t}_{1}}\rightarrow b\ell {\tilde{\nu}}$ here; they
are treated in \cite{Datta,Regina}.

When $m_{{\tilde{t}_{1}}}<m_{b}+m_{W}+m_{{\widetilde{Z}_{1}}}$, the
remaining 3-body decay is closed, and ${\tilde{t}_{1}}$ must decay as ${%
\tilde{t}_{1}}\rightarrow c{\widetilde{Z}_{1}}$ or ${\tilde{t}_{1}}%
\rightarrow b\bar{f}f^{\prime }{\widetilde{Z}_{1}}$. The rate for the loop
decay ${\tilde{t}_{1}}\rightarrow c{\widetilde{Z}_{1}}$ was estimated in 
\cite{HikKob}, and\ earlier work\cite{BDGGT,BST} had focused on the this
decay. Recent calculations\cite{Djouadi} indicate a substantial branching
fraction for the 4-body decay for realistic regions of the MSSM parameter
space. The event topologies for ${\tilde{t}_{1}}\rightarrow b\bar{f}%
f^{\prime }{\widetilde{Z}_{1}}$, but not kinematics, would be similar to
those for the ${\tilde{t}_{1}}\rightarrow b{\widetilde{W}_{1}}$ decay.\ It
is interesting to note that Ref.~\cite{Djouadi} finds a gradual transition
of dominance between the two decay modes, so that there are broad regions
where \emph{both} decays occur appreciably. This of course would greatly
complicate the search strategy and make detection more difficult. In this
work we only consider the loop decay ${\tilde{t}_{1}}{}^{\!\ast }{\tilde{t}%
_{1}}\rightarrow c\bar{c}{\widetilde{Z}_{1}}{\widetilde{Z}_{1}}$, whose
topology is $c$-jets + $\NEG{E}\mathstrut _{\top }^{\text{ }}$. The classic $%
\NEG{E}\mathstrut _{\top }^{\text{ }}+$ jets signature suffers from big
backgrounds from $W$ and $Z$\ events with jets from QCD radiation and $\NEG%
{E}\mathstrut _{\top }^{\text{ }}$ supplied by neutrinos. We will see below
that charm tagging is necessary to develop good detection schemes. Chapter 
\ref{chap:scz} is devoted to searches in this channel.

\begin{table}[tb] \centering%
$
\begin{tabular}{|c|l|}
\hline
$c{\widetilde{Z}_{1}}\mathstrut ^{\mathstrut ^{\mathstrut }}$ & discussed
fully in Chapter \ref{chap:scz} \\ 
$b{\widetilde{W}_{1}}\mathstrut _{\mathstrut _{\mathstrut }}$ & discussed
fully in Chapter \ref{chap:sbw} \\ \hline
$bW{\widetilde{Z}_{1}}\mathstrut ^{\mathstrut ^{\mathstrut }}$ & competes
with ${\tilde{t}_{1}}\rightarrow c{\widetilde{Z}_{1}}$ when $m_{{\tilde{t}%
_{1}}}\gg 165$ GeV; see \cite{porodWohr,porod} \\ 
$b\tilde{\ell}\nu ,b\ell \tilde{\nu}$ & requires light slepton; not studied
here; see \cite{Datta,Regina} \\ 
$bf\bar{f}^{\prime }{\widetilde{Z}_{1}}\mathstrut _{\mathstrut _{\mathstrut
}}$ & may compete with ${\tilde{t}_{1}}\rightarrow c{\widetilde{Z}_{1}}$;
see \cite{Djouadi} \\ \hline
\multicolumn{1}{|l|}{$c\tilde{g}\mathstrut ^{\mathstrut ^{\mathstrut }}$} & 
needs $m_{{\tilde{t}_{1}}}\gg 180$ GeV and loop-suppressed \\ 
\multicolumn{1}{|l|}{$t{\widetilde{Z}_{1}}\mathstrut _{\mathstrut
_{\mathstrut }}$} & needs $m_{{\tilde{t}_{1}}}\gg $ 210 GeV \\ \hline
\end{tabular}
$\caption{Light stop decay modes.\label{tab:decays}}%
\end{table}%

In case the ${\tilde{t}_{1}}$-${\widetilde{Z}_{1}}$ mass difference is great
enough to open up the 3-body decay, then ${\tilde{t}_{1}}\rightarrow c{%
\widetilde{Z}_{1}}$ and ${\tilde{t}_{1}}\rightarrow bW{\widetilde{Z}_{1}}$
will compete --- the 2-body decay is loop suppressed and the 3-body decay is
suppressed by phase space\cite{porodWohr,porod}. Note that for the common
ansatz $m_{{\widetilde{W}_{1}}}=2m_{{\widetilde{Z}_{1}}}$ discussed earlier,
the 3-body decay only opens up for $m_{{\tilde{t}_{1}}}>165$ GeV\ (with $m_{{%
\widetilde{W}_{1}}}>$ 160 GeV), and even then is strongly suppressed by
phase space so this mode can only be important for $m_{{\tilde{t}_{1}}}\gg
165$ GeV. This region is difficult for either the ${\tilde{t}_{1}}%
\rightarrow c{\widetilde{Z}_{1}}$ or ${\tilde{t}_{1}}\rightarrow b{%
\widetilde{W}_{1}}$ search strategies we develop below for 2 fb$^{-1}$
integrated luminosity at the Tevatron, and a direct search for ${\tilde{t}%
_{1}}\rightarrow bW{\widetilde{Z}_{1}}$ would be even more problematical due
to the presence of on-shell $W$s (whose rejection is key to suppressing SM
backgrounds). Our preliminary analysis has indicated that this decay mode is
inaccessible to the Tevatron experiments, and other investigators\cite
{porodWohr,Regina} have since found the same result.\ We do not pursue the
3-body decay further.

\miscplot{expt}{115 263 475 617}%
{Map of decay modes and experimental results for the light stop.}{%
Map of decay modes and experimental results for the light stop.
The vertical axis on the left is neutralino mass, and that on the right
is chargino mass.  In case $m_{\widetilde{W}_1}=2m_{\widetilde{Z}_1}$,
these axes coincide as illustrated. 
The heavy dotted line is the CDF 95\% exclusion contour for the
mode $\tilde{t}_1\to c\widetilde{Z}_1$.
The dashed line shows the LEP II
95\% confidence limit for neutralino mass and stop mass given the decay 
$\tilde{t}_1\to c\widetilde{Z}_1$.  The dot-dashed line is the LEP II
chargino mass limit.
}
Figure \ref{fig:expt} shows the regions where these various decays are
expected to occur. The neutralino decay mode ${\tilde{t}_{1}}\rightarrow c{%
\widetilde{Z}_{1}}$ and the chargino mode ${\tilde{t}_{1}}\rightarrow b{%
\widetilde{W}_{1}}$ are overlaid in this figure. The two axes will coincide
if $m_{{\widetilde{W}_{1}}}=2m_{{\widetilde{Z}_{1}}}$. The upper left region
marked $m_{{\tilde{t}_{1}}}<m_{{\widetilde{Z}_{1}}}$ is ruled out because ${%
\widetilde{Z}_{1}}$ is the LSP by assumption.

We restrict our attention here to physics at the Fermilab Tevatron $p\bar{p}$
collider (described in Section 1.3.1). There also exists a rich stop
phenomenology at $e^{+}e^{-}$ colliders (see \cite{Bartl} and references
therein), and proposed $\mu ^{+}\mu ^{-}$ colliders\cite{mumu}. We do not
examine $R$-parity violating processes for the Tevatron, which are discussed
in \cite{deCampos} and \cite{Berger}.

\subsubsection{A note on the top sector}

Since we are interested in assessing the reach for the SUSY signal over
Standard Model backgrounds, we assume that the top always decays via $%
t\rightarrow bW$ and neglect SUSY decays of tops such as $t\rightarrow {%
\tilde{t}_{1}\widetilde{Z}_{1}}$\cite{Hosch,LiOaks,Sender} or $t\rightarrow {%
\tilde{t}_{1}\tilde{g}}$\cite{ShanZhu}. If these decays are present to any
extent, they could potentially add to the signal as the stops cascade via
the decays discussed above. That is, a $t\bar{t}$ event where one of the
tops decays supersymmetrically will be less likely to fail the cuts designed
to reject the $t\bar{t}$ background, and so will improve the yield.\ Our
assumption leads to a conservative background estimate, and if anything will
underestimate the discovery potential for stops.

\section{Current experimental results}

There are good experimental limits on the stop mass from both the CERN\ LEP\
II\ $e^{+}e^{-}$ collider and the Fermilab Tevatron $p\bar{p}$ collider.
LEP\ II\ results\cite{OPALstop,L3stop,ALEPHspart} at a collision energy $%
\sqrt{s}=189$ GeV set a 95\% confidence limit exclusion on stops lighter
than 90.3 GeV for the decay mode ${\tilde{t}_{1}}\rightarrow c{\widetilde{Z}%
_{1}}$, assuming the stop is at least 10 GeV heavier than the neutralino.
This limit weakens by a few GeV when $\theta _{t}=0.98$ and ${\tilde{t}_{1}}$
decouples from the $Z$. The LEP experiments do not report a stop mass limit
for the decay ${\tilde{t}_{1}}\rightarrow b{\widetilde{W}_{1}}$, but their
chargino mass limit (see below)\ implies $m_{{\tilde{t}_{1}}}>98$ GeV if
stops decay via this mode.

\looseness=-2
At the Tevatron, the D0 collaboration\cite{D0metjet} investigated the ${%
\tilde{t}_{1}}\rightarrow c{\widetilde{Z}_{1}}$ mode in the $\NEG%
{E}\mathstrut _{\top }^{\text{ }}$ + jets channel (see Fig.~\ref{fig:expt} and
Section 2.2) and excluded the region bounded by $m_{{\tilde{t}_{1}}}<100$\
GeV, $m_{{\widetilde{Z}_{1}}}<$ 40 GeV and $m_{{\tilde{t}_{1}}}-m_{{%
\widetilde{Z}_{1}}}>$ 40 GeV. The CDF collaboration\cite{CDFscz} got a
bigger excluded region by analyzing 88 pb$^{-1}$ of Run I data (see Chapter 
\ref{chap:tech}) in the $\NEG{E}\mathstrut _{\top }^{\text{ }}+$ jet + charm
jet channel for this mode and set a 95\% CL exclusion for a region roughly
bounded by $m_{{\tilde{t}_{1}}}<115$ GeV and $m_{{\widetilde{Z}_{1}}}<50$
GeV. The Tevatron collaborations also searched for the ${\tilde{t}_{1}}%
\rightarrow b{\widetilde{W}_{1}}$ mode. CDF looked for a $\NEG{E}\mathstrut
_{\top }^{\text{ }}$ + lepton + jets (from $b$-quarks)\cite{CDFsbw}, and D0
looked in the dielectron channel\cite{D0sbw}. Neither group was able to
improve the existing limits (of $m_{{\tilde{t}_{1}}}>m_{{\widetilde{W}_{1}}%
}+m_{b}>95$ GeV).

LEP\ II\ also provides the best limits on the ${\widetilde{W}_{1}}$ and ${%
\widetilde{Z}_{1}}$ masses. Chargino limits are generally function of $m_{0}$%
; small $m_{0}$ allows ${\widetilde{W}_{1}}$ and ${\widetilde{Z}_{1}}$ to be
nearly degenerate in mass which seriously compromises the detection
efficiency and hence the mass reach. The latest published results (for a
collision energy $\sqrt{s}\leq 189$ GeV)\cite{OPALino,L3ino,ALEPHino} find $%
m_{{\widetilde{W}_{1}}}>$ 93 GeV and $m_{{\widetilde{Z}_{1}}}>32$ GeV for $%
m_{{\widetilde{W}_{1}}}\gg m_{{\widetilde{Z}_{1}}}$ (weakening to $m_{{%
\widetilde{W}_{1}}}>72$ GeV, $m_{{\widetilde{Z}_{1}}}>31$ GeV in the worst
case).

Experiments at LEP\ II\ are ongoing, and limits from there are evolving even
as this thesis is being written. The experiments now have accumulated data
at $\sqrt{s}=200$ GeV which will published shortly. For instance, ALEPH\cite
{ALEPH200} has just released preliminary bounds of $m_{{\widetilde{W}_{1}}%
}>100$ GeV, $m_{{\widetilde{Z}_{1}}}>35$ GeV, $m_{\tilde{e}}>92$ GeV, $m_{%
\tilde{\mu}}>85$ GeV, $m_{H}>92$ GeV, and $\tan \beta >1.9$.

\chapter{Tevatron Simulation%
\label{chap:tech}%
}

\section{Fermilab Tevatron}

The Fermilab Tevatron is currently the world's premier (indeed, only) $p\bar{%
p}$ collider. It has two major general purpose collider detectors, CDF\cite
{CDFdetector,CDFdetector2} and D0\cite{D0detector}\cite{D0detector2}. For
reference, we group Tevatron operations into three phases: Run I, Run II and
Run II+. Run I refers to the data already collected during the machine's
previous operating phase in 1994-1995. Run I obtained an integrated
luminosity of almost 100 $\mathrm{pb}^{-1}$ in each detector\ at a
center-of-mass energy of $\sqrt{s}=$ 1.8 TeV. These data have already been
analyzed for signatures of the light stop, and the results of these analyses
are presented below. Table \ref{tab:runs} lists the parameters for these
runs \cite{Weerts}.

\begin{table}[b] \centering%
$
\begin{tabular}{l|c|c|c|c|}
\cline{2-2}\cline{2-5}\cline{4-4}
& $\sqrt{s}$ (TeV) & $\mathcal{L}$ (cm$^{-2}$sec$^{-1}$) & $\int \mathcal{L}%
dt$ (fb$^{-1}$) & dates \\ \hline
\multicolumn{1}{|l|}{Run I} & 1.8 & $\leq 10^{31}$ & $\sim 0.1$ & 1994-1995
\\ 
\multicolumn{1}{|l|}{Run II} & 2.0 & $2\times 10^{32}$ & $\sim 2$ & 2001-2003
\\ 
\multicolumn{1}{|l|}{Run II+} & 2.0 & $2-5\times 10^{32}$ & $>15$ & to 2007
\\ \hline
\end{tabular}
$\caption{Tevatron run parameters.\label{tab:runs}}%
\end{table}%

The Tevatron is currently undergoing major machine and detector upgrades and
is scheduled to resume operation in 2001. Chief among the upgrades is the
new ``Main Injector'' which will store and pre-accelerate protons and
antiprotons for injection into the Tevatron ring itself. The Main Injector
will also recycle antiprotons from the Tevatron resulting in a tenfold
increase in luminosity. Together with the new Recycler Ring, the peak
luminosity is expected to rise by a factor 20 from $1\times 10^{31}$ cm$%
^{-2} $sec$^{-1}$ to $2\times 10^{32}$ cm$^{-2}$sec$^{-1}$. Additionally,
the beam energy will be raised to $\sqrt{s}=$ 2.0 TeV (which will increase
the $\tilde{t}^{\ast }\tilde{t}$ and $t\bar{t}$ production cross-sections by
40\% over those at $\sqrt{s}=$ 1.8 TeV). The primary data collection period
following the Main Injector upgrade is referred to as Run II. During Run II
we expect that 2 $\mathrm{fb}^{-1}$\ of data will be accumulated at 2.0 TeV
after about two years of operation.

The Tevatron is expected to continue operations for several years beyond the
nominal 2 fb$^{-1}$ Run II phase. There are also proposals on the table for
further major luminosity upgrades.\ In anticipation of a substantially
larger data sample, we also make projections for an integrated luminosity of
25 $\mathrm{fb}^{-1}$\ at 2.0 TeV, which we call Run II+.

\section{Event Simulation}

Collisions at $p\bar{p}$ machines are messy affairs\cite{BargerPhillips}. At
2 TeV the proton is a complicated assemblage of quark and gluon ``partons''
interacting via the strong force. In a hard-scattering event almost always
just one parton from the proton and one parton from the antiproton interact.
Each scattering parton has a certain probability to be of a given type (say
gluon, strange quark, or whatever) and to carry a given fraction of the
hadron's momentum. The set of functions which encode these probabilities for
given momentum transfer in the hard-scattering sub-process are called parton
distribution functions. They are fundamental inputs in any model that
describes large momentum transfer hadron scattering processes..

The partons can be treated as free on their way to the hard-scattering, as
are their colored reaction products. While inbound they radiate gluons
freely, some of which are hard enough to be resolved as initial-state
radiation (ISR). The partons being free at the collision point, the hard
scattering reaction itself can be calculated in perturbation theory. After
the collision the reaction products recoil from each other and the ISR.
Color confinement causes the colored reaction products, any final-state
radiation, the ISR, and the remnants of the initial $p$ and $\bar{p}$ to
hadronize as they flee the collision point. The usual picture is of color
flux lines which stretch and break in a process called fragmentation. Each
escaping colored particle will evolve into a multitude of color singlet
hadrons, all moving in rough collinearity to the original particle. This
concentrated shower of particles is called a ``jet.'' (The remnants of the
original $p$ and $\bar{p}$ move down the beam axis to form ``beam jets.'')

The hard scattering processes are described by the underlying high energy
theory. In our case , these are 
\begin{eqnarray}
\begin{array}{c}
q\bar{q}\rightarrow {\tilde{t}_{1}^{\ast }\tilde{t}_{1}} \\ 
gg\rightarrow {\tilde{t}_{1}^{\ast }\tilde{t}_{1}}
\end{array}
\end{eqnarray}
and are described by the supersymmetric model discussed in Chapter 1. The
decays of the stops into SM quarks are also described by this theory. The
processes by which the partons fragment and ultimately produce hadronic
jets, leptons, photons and so on that are detected in the experimental
apparatus is independent of the high energy theory, and phenomenological
models that describe these are encoded in event generators that we outline
below.

The (stable) hadrons, charged leptons, or photons produced in the collision
move away from the collision point and record their passage in the
surrounding detector. The detector is a complicated apparatus designed to
characterize the visible reaction products. It has components for tracking
charged particles, calorimetry for absorbing and measuring the energy
carried away from the collision by the hadrons as well as electromagnetic
particles (mainly electrons and photons), and on the outside a system for
measuring muon momenta. Modern detectors also include a sophisticated system
that serves to resolve displaced secondary vertices from the decay of heavy
flavors.

The fraction of the hadrons' longitudinal momentum carried by the initial
hard scattering partons is unknowable, so there is an irreducible
uncertainty in the longitudinal boost of the center-of-momentum frame for
the colliding system. However, by forming the vector sum of all the energy
deposited in the transverse direction, we get an important quantity called
missing transverse energy, $\NEG{E}\mathstrut _{\top }^{\text{ }}$. A
certain amount of $\NEG{E}\mathstrut _{\top }^{\text{ }}$ is due to jet and
lepton mismeasurement from imperfect energy resolution, particles going into
cracks in the detector, and other ``non-physics'' causes. A large $\NEG%
{E}\mathstrut _{\top }^{\text{ }}$, however, generally indicates the
production of one or more high-energy long-lived weakly-interacting
particles that escape the experimental apparatus without depositing energy.
In the Standard Model these would be neutrinos. In SUSY searches, large $\NEG%
{E}\mathstrut _{\top }^{\text{ }}$ is the signature of escaping LSPs.
Indeed, since a pair of LSPs are always produced in a SUSY reaction (for $R$%
-parity a good symmetry), large $\NEG{E}\mathstrut _{\top }^{\text{ }}$ is
the hallmark of a supersymmetric reaction.

\subsection{Collision simulation}

We use the program {\small ISAJET}\ v7.29\cite{isajet2} with the {\small %
ISASUSY}\cite{isasusy} extension to simulate Tevatron events for both ${%
\tilde{t}_{1}}{}^{\!\ast }{\tilde{t}_{1}}$ and background processes. {\small %
ISAJET}\ does Monte Carlo event generation for $p\bar{p}$ (and other)\
collisions. It supports a wide variety of parton distribution function sets
(We use the {\small CTEQ2L} set.).\ It incorporates lowest order
perturbative QCD cross-sections, initial- and final-state QCD radiation,
independent fragmentation and final-state hadronization, and underlying
event effects. It includes a phenomenological model tuned to minimum bias
and hard scattering data for the beam jets. {\small ISASUSY} provides
supersymmetric particle production and branching ratio calculations, and
keeps track of all SUSY decay chains. A {\small SUGRA} package accepts
GUT-scale input (Eq.~(\ref{eq:sugra:params})) and applies the RGE to produce
low-energy MSSM parameters relevant for phenomenological calculations.

\looseness=2
Because we control the relevant SUSY parameter space directly, we use the 
{\small ISASUSY} \texttt{MSSM} technique to set SUSY masses $m_{{\tilde{t}%
_{1}}}$, $m_{{\widetilde{Z}_{1}}}$ and $m_{{\widetilde{W}_{1}}}$ by hand.
Events are generated with the reaction products' transverse momentum $%
p_{\top }$ ranging from 20-320 GeV (corresponding to {\small ISAJET}'s 
\texttt{PT} and \texttt{QTW} cards). The Monte Carlo routine inefficiently
samples over large $p_{\top }$ ranges, so each calculation is automatically
broken into subranges of 20-40, 40-80, 80-160 and 160-320 GeV, and the
results combined. We typically generate 50 fb$^{-1}$ of sample for each
signal and background case.

\subsection{Detector simulation}

As each event is generated, it is processed through a toy detector simulator
consisting of the following elements (where $\eta $ is the pseudorapidity, $%
\phi $ is the azimuthal angle, and $\Delta R=\sqrt{(\Delta \eta
)^{2}+(\Delta \phi )^{2}}$).

\begin{itemize}
\item  Calorimeter simulator: We implement a toy calorimeter based on the 
{\small ISAJET}\ \texttt{CALSIM} subroutine. The segmentation is $\Delta
\eta \times \Delta \phi =0.1\times 0.087$ extending to a rapidity of $|\eta
|=4$. There is a hadronic calorimeter, into which hadrons deposit their
energy with a resolution of $50\%/\sqrt{E_{\mathrm{\top }}^{{}}}$, and an
electromagnetic calorimeter which captures electrons and photons with
resolution $15\%/\sqrt{E_{\mathrm{\top }}^{{}}}$. We do not attempt to
simulate effects of cracks or dead regions that are specific to particular
detectors.

\item  Isolated lepton identification: We sum the hadronic transverse energy
in a cone of $\Delta R<0.4$ around each lepton. If this hadronic energy is
less than 25\% of the lepton's transverse energy, then the lepton is
declared isolated. $p\mathstrut _{\top }^{\text{ }}$ thresholds for isolated
leptons are given for the relevant channels.

\item  Jet identification: We use {\small ISAJET}'s \texttt{GETJET}
jet-finding algorithm. Jets are defined as hadronic clusters with total $E_{%
\mathrm{\top }}^{{}}>$ 15 GeV falling within a cone of radius $\Delta R<0.7$
and subject to $|\eta |<3.5$. We do not perform jet energy corrections.

\item  Silicon vertex detector (SVX): We simulate a SVX detector for tagging 
$b$-jets. We identify each weakly-decaying $B$ hadron in an event with $E_{%
\mathrm{\top }}^{{}}>15\mathrm{\ GeV}$ and $|\eta |<2$. If $\Delta R(B,\text{%
jet})<0.5$ for some jet then that jet is tagged as a $b$-jet with an
efficiency of 50\%.

\item  Charge multiplicity counter: The charge multiplicity of a jet is
defined as the number of charged long-lived hadrons with $E_{\mathrm{\top }%
}^{{}}>0.5\mathrm{\ GeV}$ lying within $\Delta R<0.5$ of the jet axis. We
use this for $\tau $ veto studies.

\item  Top mass reconstruction: For single-lepton events we perform $t\bar{t}
$ event reconstruction and calculate the resulting top mass. Details of this
procedure are given in Chapter \ref{chap:sbw}.
\end{itemize}

Events are produced by {\small ISAJET}, passed through this detector
simulator and subjected to software ``trigger'' cuts. Those which pass these
cuts have their event data written. This data is then post-processed to
prepare ntuple files for later cut analysis using the CERN Physics Analysis
Workstation (PAW) software\cite{PAW}.

\section{Methodology}

We study the decay modes ${\tilde{t}_{1}}\rightarrow b{\widetilde{W}_{1}}$
and ${\tilde{t}_{1}}\rightarrow c{\widetilde{Z}_{1}}$ separately, in each
case taking the ${\tilde{t}_{1}}$ branching fraction to be 100\% for the
given mode. For the ${\tilde{t}_{1}}\rightarrow c{\widetilde{Z}_{1}}$ mode,
the stop signal then depends only on the two SUSY masses $m_{{\tilde{t}_{1}}%
} $ and $m_{{\widetilde{Z}_{1}}}$, and is independent of mixing angles in
the top squark and gaugino sectors.

On the other hand, signals from ${\tilde{t}_{1}}\rightarrow b{\widetilde{W}%
_{1}}$ mode events (with the cascade decay ${\widetilde{W}_{1}}\rightarrow {%
\widetilde{Z}_{1}}+\text{fermions}$) are determined by the three SUSY masses 
$m_{{\tilde{t}_{1}}}$, $m_{{\widetilde{W}_{1}}}$ and $m_{{\widetilde{Z}_{1}}%
} $, together with the branching fractions for ${\widetilde{W}_{1}}%
\rightarrow e,\mu ,\tau $. In most of our analysis, we collapse this
parameter space from four dimensions to two. First, we generally take $m_{{%
\widetilde{Z}_{1}}}=m_{{\widetilde{W}_{1}}}/2$, which is approximately true
in many interesting classes of SUSY models (see eqn.~(\ref{eq:sugra:winozino}%
)). This only requires that the lighter chargino and the two lighter
neutralino are gaugino-like, and that the gaugino masses obey the mass
unification condition. While this occurs canonically in grand unified
models, there are also many other models where this condition is obeyed but
for entirely different reasons. Second, we set the chargino's leptonic
branching fraction equal to the $W$'s leptonic branching fraction: 
\begin{eqnarray}
\mathcal{B}({\widetilde{W}_{1}}\rightarrow e)=\mathcal{B}({\widetilde{W}_{1}}%
\rightarrow \mu )=\mathcal{B}({\widetilde{W}_{1}}\rightarrow \tau )=\mathcal{%
B}(W\rightarrow e)\approx 1/9, 
\end{eqnarray}
as discussed earlier. We also examine how our conclusions are affected when
these restrictions are relaxed.

For each channel investigated, we generate event data sets for a grid over
the appropriate parameter space of relevant SUSY masses, together with event
data sets for the relevant background processes. The grid spacing is usually 
$\Delta m_{{\tilde{t}_{1}}}=10$ GeV, $\Delta m_{{\widetilde{W}_{1}}}=10$ GeV
and $\Delta m_{{\widetilde{Z}_{1}}}=5$ GeV. These data sets are then
subjected to a variety of cuts on measurable quantities designed to maximize
the discovery reach over the SUSY parameter space, using the PAW software to
analyze cuts.

We consider a particular SUSY model ``discoverable'' if (1) there is a $%
5\sigma $ statistical significance 
\begin{eqnarray}
\frac{N_{s}}{\sqrt{N_{b}}}>5,%
\label{eq:5sigma}%
\end{eqnarray}
where $N_{s}$ $\left( N_{b}\right) $ is the expected number of signal $%
\left( \text{background}\right) $ events, (2) the expected signal level is
at least 25\% of background, 
\begin{eqnarray}
\frac{N_{s}}{N_{b}}>25\%,%
\label{eq:25pct}%
\end{eqnarray}
and (3)\ at least 5 signal events are expected 
\begin{eqnarray}
N_{s}\geq 5. 
\end{eqnarray}
When comparing our results to those of other studies, bear in mind that many
authors report their results at a $3\sigma $ level rather than our $5\sigma $
level. The 25\% requirement eqn.~(\ref{eq:25pct}) encodes our overall
uncertainty in the calculated absolute cross-sections for background
processes. And we require that at least 5 signal events be expected for
protection against uncontrolled non-physics backgrounds.

A signal which fails $N_{s}/N_{b}>25\%$ is called ``background limited.''
Increasing the integrated luminosity $\widehat{\mathcal{L}}$ cannot help
this situation, and only clever cuts can rescue the signal. When a signal
fails $N_{s}/\sqrt{N_{b}}>5$ it is ``rate limited.'' A rate-limited\ signal
that just satisfies $N_{s}/N_{b}=25\%$ can be recovered by increasing $%
\widehat{\mathcal{L}}$ until the point $N_{s}/\sqrt{N_{b}}=5\sigma $ is
reached. At this point, where the equalities in (\ref{eq:5sigma}) and (\ref
{eq:25pct}) both hold, we have $N_{B}=400$ and $N_{S}=100$. Writing $%
N_{B}=\sigma _{B}\widehat{\mathcal{L}}$, with $\sigma _{B}$ the background
cross-section, then 
\begin{eqnarray}
\widehat{\mathcal{L}}=\frac{400}{\sigma _{B}^{{}}}\text{\qquad when }\frac{%
N_{s}}{\sqrt{N_{b}}}=5\sigma \text{ and }\frac{N_{s}}{N_{b}}=25\%%
\label{eq:400}%
\end{eqnarray}
which will be useful to us in the sequel.

The observability criteria adopted here are conservative, and elaborate
statistical schemes might well extend our reported reach considerably. We
have, for instance, not investigated the utility of using information from
the background-rich kinematical regions excluded by our cuts to get a better
data-derived background estimation which would allow us to relax our 25\%
criterion. Also, it is worth remembering that by combining results from both
CDF and D0 one can have get a stronger signal than either detector would get
alone.

\chapter{The decay mode ${\widetilde{%
\lowercase{t}%
}_{1}}\rightarrow {%
\lowercase{c}%
}{\widetilde{Z}_{1}}$%
\label{chap:scz}%
}

If the top squark decays via ${\tilde{t}_{1}}\rightarrow c{\widetilde{Z}_{1}}
$, it can be searched for in the multijet + $/\!\!\!\!E_{\mathrm{T}}$
channel, which is the standard channel for SUSY searches at hadron
colliders. This channel picks out events where the two ${\widetilde{Z}_{1}}$
neutralinos escape detection in the experimental apparatus, resulting in a
large net missing transverse energy. We also investigate the possibility of
tagging one of the charm jets via its muon decay to further enhance the
signal over SM backgrounds. We take the branching fraction for ${\tilde{t}%
_{1}}\rightarrow c{\widetilde{Z}_{1}}$ to be 100\%. The SUSY signal in each
of these channels is then completely determined by the two quantities $m_{{%
\tilde{t}_{1}}}$ and $m_{{\widetilde{Z}_{1}}}$. The stop mass $m_{{\tilde{t}%
_{1}}}$ sets the production cross section and the two masses together
determine the kinematics. We use {\small ISAJET}\ to simulate signal events
over a grid in the $m_{{\tilde{t}_{1}}}-m_{{\widetilde{Z}_{1}}}$ plane. For
each value of $(m_{{\tilde{t}_{1}}},m_{{\widetilde{Z}_{1}}})$ we generate
events, run them through our toy detector simulation and compare the
resulting cross-sections with those from the SM background processes to
determine the reach of the Tevatron experiments.

\section{Missing $E_{\top }$ + jets\ channel}

\miscplot{met-topo}{140 566 302 682}%
{Event topology for the \metjet\ channel}{%
The standard event topology for the \metjet\ channel of the 
$\scz$ decay mode.
}

The ${\tilde{t}_{1}^{\ast }\tilde{t}_{1}}$ event topology is exhibited in
Fig.~\ref{fig:met-topo}. The two LSPs combine to provide a large $\NEG%
{E}\mathstrut _{\top }^{\text{ }}$, and the two charm quarks are required to
provide hard jets. Standard Model events that have hard neutrinos (to supply 
$\NEG{E}\mathstrut _{\top }^{\text{ }}$) without isolated leptons, as in
Fig.~\ref{fig:met-zbkg}, are the main physics backgrounds to this ${\tilde{t}%
_{1}}\rightarrow c{\widetilde{Z}_{1}}$ search.\ $Z\rightarrow \nu \bar{\nu}$%
\ is a problem when direct high $p\mathstrut _{\top }^{\text{ }}$ weak
boson\ production is accompanied by gluon or quark radiation, and also when
an outgoing quark from a QCD hard-scattering radiates a $Z$ which decays
invisibly. We use jet hardness and geometrical cuts to suppress these
relative to the signal.\ Similarly, events with $W\rightarrow \tau \nu $\
contribute to the background when the $\tau $ decays hadronically or into a
non-isolated lepton and fails to be distinguished from a hadronic jet. The\
processes of concern are $q\bar{q}^{\prime }\rightarrow W\rightarrow \tau
\nu $ plus one or two QCD jets and $gq\rightarrow Wq^{\prime }\rightarrow
\tau \nu q^{\prime }$. $W$s from $t\bar{t}$ events also contribute to the
background.

\miscplotb{met-zbkg}{136 564 498 682}%
{Background processes for the \metjet\ channel}{%
Some representative background processes for the \metjet\ channel.
}

Note that within {\small ISAJET}, $Z$-production processes such as $%
gq\rightarrow qZ$ with final-state gluon radiation (Fig.~\ref{fig:met-zkinds}%
a) are treated separately from QCD hard-scattering processes such as $%
gq\rightarrow gq$ or $q^{\prime }q\rightarrow q^{\prime }q$ with final-state 
$Z$ radiation (Fig.~\ref{fig:met-zkinds}b). While the former are readily and
efficiently simulated with the {\small ISAJET} \texttt{DRELLYAN} reaction
card, the latter are treated by {\small ISAJET} as QCD processes and it is
difficult to obtain a large Monte Carlo integrated luminosity for the latter
due to the huge QCD hard scattering cross section. By analyzing $\sim 10^{7}$
events, we found that these radiative events add about 5\% to the related $q%
\bar{q}\rightarrow gZ$ background.

\miscplot{met-zkinds}{146 579 386 680}%
{{\small ISAJET} $Z$ radiative processes}{%
Examples of {\small ISAJET} $Z$ radiative processes. Diagram (a)
is a well-controlled {\small ISAJET} {\texttt{DRELLYAN}} process; 
processes like (b) require simulation of a very large number of
{\small ISAJET} {\texttt{QCD}} events.
}

\subsection{Run I%
\label{sec:met:runI}%
}

We performed an early analysis\cite{BST} to estimate discovery prospects in
this channel with the Tevatron Run I data. Based on this analysis we
recommended a set of cuts: (1)\ $\NEG{E}\mathstrut _{\top }^{\text{ }}{>50%
\mathrm{\ GeV}}$ to reduce backgrounds from QCD heavy flavors and
mismeasured jets; (2)\ at least two jets in the event, one of which is
central (${|\eta |<1)}$; (3) all jets separated by at least 30$^{\text{o}}$
in azimuth from $\NEG{E}\mathstrut _{\top }^{\text{ }}$, and a jet-jet
separation less than 150$^{\text{o}}$ in case there are only two jets; and
(3)\ no identified leptons. We also prescribed a compound cut (4)\ ${%
p\mathstrut _{\top }^{\text{ }}}\left( \text{fast jet}\right) >80$ GeV for $%
\Delta \phi ({\NEG{E}\mathstrut _{\top }^{\text{ }},j)>90}^{\circ }$, else ${%
p\mathstrut _{\top }^{\text{ }}}\left( \text{fast jet}\right) >50$ GeV. We
found the dominant background after these cuts to be $W\rightarrow \tau \nu $%
, with the $\tau $ decaying hadronically. Depending on the ability of
experimentalists to reject the $W\rightarrow \tau \nu $ background and
subtract the $Z\rightarrow \nu \bar{\nu}$ backgrounds, we estimated that,
depending on $m_{{\widetilde{Z}_{1}}}$, stops as heavy as 100-125 GeV might
be probed in this channel with 100 pb$^{-1}$ of data.

So far, the full Run I data set has not been analyzed in this channel. The
D0 Collaboration\cite{D0metjet} looked at 13.5 pb$^{-1}$ of data from the
1992-1993 run. They required $\NEG{E}\mathstrut _{\top }^{\text{ }}{>40%
\mathrm{\ GeV}}$, two jets with ${p\mathstrut _{\top }^{\text{ }}>30\mathrm{%
\ GeV}}$, $90^{\circ }<{\Delta \phi (j_{1},j_{2})<165^{\circ }}$, $10^{\circ
}<{\Delta \phi (\NEG{E}\mathstrut _{\top }^{\text{ }},j)}$, ${\Delta \phi (%
\NEG{E}\mathstrut _{\top }^{\text{ }},j)<125}^{\circ }$ and no identified
leptons with ${p\mathstrut _{\top }^{\text{ }}>10\mathrm{\ GeV.}}$ They
excluded a region at the 95\% confidence level of $m_{{\tilde{t}_{1}}%
}\lesssim 90$ GeV, $m_{{\tilde{t}_{1}}}-m_{{\widetilde{Z}_{1}}}\gtrsim 35$
GeV.

\subsection{Run II}

The larger data sample anticipated in Run II enables us to sharpen the
selection cuts for the stop signal. For the $\NEG{E}\mathstrut _{\top }^{%
\text{ }}$ + jets\ channel we require the following initial cuts:%
\begin{eqnarray*}
\left( i\right) &&\NEG{E}\mathstrut _{\top }^{\text{ }}{>50\mathrm{\ GeV};}
\\
\left( ii\right) &&\text{at least }{\text{two jets with }p\mathstrut _{\top
}^{\text{ }}>50\mathrm{\ GeV},\text{ one with }|\eta |<1;} \\
\left( iii\right) &&{\text{for all jets, }\Delta \phi (\NEG{E}\mathstrut
_{\top }^{\text{ }},j)>30^{\circ },} \\
\quad &&\quad \quad \quad \quad {\text{and if only two jets then }\Delta
\phi (j_{1},j_{2})<150^{\circ };} \\
\left( iv\right) &&{\text{no identified isolated }}e\text{ or }\mu {\text{;}}
\\
\left( v\right) &&{\text{no SVX }}b\text{ }{\text{tag.}}
\end{eqnarray*}
These cuts will reject QCD backgrounds from heavy flavor production ($%
g\rightarrow c\bar{c},b\bar{b})$ and multijet production with mismeasured
jets. Cuts $\left( i\right) $\ or $\left( ii\right) $\ may serve as
triggers. Note that we require very hard jets, one of which must be central.
Cut $\left( iii\right) $\ is designed to reduce detector-dependent
backgrounds due to mismeasured QCD jets. Cut $\left( iv\right) $\ suppresses
background events with leptons from $W\rightarrow \ell \nu $, $\ell =e,\mu $%
. Cut $\left( v\right) $\ suppresses the $t\bar{t}$\ background by vetoing
events with $b$-jets which will be efficiently tagged by the Run II Tevatron
detectors. (In our simulation we ignore $c$ contamination of SVX tags. A
closer analysis which pays attention to the detailed experimental properties
of the particular vertex detector, including a scheme for the outright
identification of displaced charm vertices, could improve the signal yield
in this channel. See the remarks at the end of the next section.) The cross
sections for the backgrounds and selected SUSY signal cases after these cuts
are displayed in Table \ref{tab:met} in the column labelled $\left( v\right) 
$.%
\cutgroup{met}
\begin{table} \centering%
\begin{tabular}{|c|c|c|c|c|c|}
\hline
background & $\left( v\right) $ & $\left( vi\right) $ & $\left( vii\right) $
& $\left( viii\right) $ & $\left( viii'\right) $ \\ \hline
$W\rightarrow \tau \nu $ & \multicolumn{1}{|r|}{500} & \multicolumn{1}{|r|}{
391} & \multicolumn{1}{|r|}{246} & \multicolumn{1}{|r|}{69} & 
\multicolumn{1}{|r|}{13.4} \\ 
$Z\rightarrow \nu \bar{\nu}$ & \multicolumn{1}{|r|}{484} & 
\multicolumn{1}{|r|}{268} & \multicolumn{1}{|r|}{157} & \multicolumn{1}{|r|}{
105} & \multicolumn{1}{|r|}{24.4} \\ 
$t\bar{t}$ (175) & \multicolumn{1}{|r|}{32} & \multicolumn{1}{|r|}{24} & 
\multicolumn{1}{|r|}{24} & \multicolumn{1}{|r|}{7} & \multicolumn{1}{|r|}{2.5
} \\ \hline
total & \multicolumn{1}{|r|}{1016} & \multicolumn{1}{|r|}{683} & 
\multicolumn{1}{|r|}{427} & \multicolumn{1}{|r|}{181} & \multicolumn{1}{|r|}{
40.3} \\ \hline
25\% & \multicolumn{1}{|r|}{254} & \multicolumn{1}{|r|}{171} & 
\multicolumn{1}{|r|}{107} & \multicolumn{1}{|r|}{45} & \multicolumn{1}{|r|}{
10.1} \\
5$\sigma $ @ 2fb$^{-1}$ & \multicolumn{1}{|r|}{113} & \multicolumn{1}{|r|}{92
} & \multicolumn{1}{|r|}{73} & \multicolumn{1}{|r|}{48} & 
\multicolumn{1}{|r|}{(6.4)} \\ \hline\hline
${\tilde{t}_{1},}$ ${\widetilde{Z}_{1}}$ & $\left( v\right) $ & $\left(
vi\right) $ & $\left( vii\right) $ & $\left( viii\right) $ & $\left(
viii'\right) $ \\ \hline
120, 80 & \multicolumn{1}{|r|}{146} & \multicolumn{1}{|r|}{117} & 
\multicolumn{1}{|r|}{77} & \multicolumn{1}{|r|}{47} & \multicolumn{1}{|r|}{
12.1} \\ 
150, 80 & \multicolumn{1}{|r|}{146} & \multicolumn{1}{|r|}{120} & 
\multicolumn{1}{|r|}{86} & \multicolumn{1}{|r|}{58} & \multicolumn{1}{|r|}{
19.7} \\ 
180, 100 & \multicolumn{1}{|r|}{68} & \multicolumn{1}{|r|}{54} & 
\multicolumn{1}{|r|}{42} & \multicolumn{1}{|r|}{29} & \multicolumn{1}{|r|}{
11.0} \\ \hline
\end{tabular}
\caption[$/\!\!\!\!E_{{\rm T}}$ + jets\ channel cross-sections]{%
$/\!\!\!\!E_{{\rm T}}$ + jets\ channel cross-sections, in fb, after cuts as
discussed in the text. The line labelled ``25\%'' shows 25\% of the total
background, and the ``$5\sigma$" line shows the signal level needed to
produce a $5\sigma $ excess at 2 $fb^{-1}$. (The parenthesized $5\sigma $
value in the last column is for an integrated luminosity of 25 $fb^{-1}$.)
For the signal cases, $\tilde{t}_{1}$ and $\widetilde{Z}_{1}$ masses are
given in GeV.} \label{tab:met}%
\end{table}%

For the $Z\rightarrow \nu \bar{\nu}$\ background due to Drell-Yan $q\bar{q}$
annihilation, the two hard jets come from QCD radiation recoiling against
the $Z$ boson; these jets are often close together in azimuth. The $c$-jets
in the signal, on the other hand, come from the two ${\tilde{t}_{1}}$s which
recoil mainly against each other, and hence tend to be more back to back in $%
\phi $. Events from the $W\rightarrow \tau \nu $\ background exhibit a
spectrum intermediate between these two; when the two hardest jets come from
QCD radiation then they are generally close, as with $Z\rightarrow \nu \bar{%
\nu}$, but when one of the hard jets is from a hadronically decaying $\tau $%
, then the jets can have a large separation. Distributions of the azimuthal
separation of the two hardest jets, $\Delta \phi (j_{1},j_{2})$, are
displayed in Fig.~\ref{fig:met-phijj} for the $W$ and $Z$\ backgrounds and a
signal case of a 150 GeV ${\tilde{t}_{1}}$ decaying to an 80 GeV ${%
\widetilde{Z}_{1}}$. Imposing the cut 
\begin{eqnarray*}
\left( vi\right) \quad {\Delta \phi (j_{1},j_{2})>90^{\circ }} 
\end{eqnarray*}
eliminates about half of the $Z\rightarrow \nu \bar{\nu}$\ background along
with 20\% of the signal and other backgrounds.%

\miscplot{met-phijj}{115 216 518 492}%
{Distribution of $\phijj$}{%
Distribution of $\phijj$ in the \metjet\ channel for a signal case of 
150 GeV $\tilde{t}_1 \to$ 80 GeV $\widetilde Z_1$ and backgrounds
\wtn\ and \znn.
The vertical axis on the main plot is in fb/bin.  The top line
in the histogram plot shows the sum of signal plus both
backgrounds.  The small subplot at the top shows the ratio of
the normalized signal distribution to the normalized signal+background
distribution.  The box heights in this subplot are proportional
to their calculated uncertainties.
}

We see from Table \ref{tab:met} that while a $5\sigma $ signal is already
possible even for $m_{{\tilde{t}_{1}}}=150\mathrm{\ GeV}$, the signal to
background ratio is less than 25\%. To enhance the signal further, we note
that these signal events generally have harder jets than the $W$ and $Z$\
events, except when the ${\tilde{t}_{1}}-{\widetilde{Z}_{1}}$ mass
difference becomes small. Also, the ${\tilde{t}_{1}^{\ast }\tilde{t}_{1}}$
events may be accompanied by QCD jets. The distribution of total jet
transverse energy 
\begin{eqnarray}
J_{\mathrm{\top }}\equiv \sum_{\mathrm{jets}}E_{\mathrm{\top }}^{\mathstrut
}\left( \text{jet}\right) 
\end{eqnarray}
exhibited in Fig.~\ref{fig:met-jt} reflects this. We make the cut 
\begin{eqnarray*}
\left( vii\right) \quad {J_{\mathrm{\top }}>175\mathrm{\ GeV,}} 
\end{eqnarray*}
which reduces the $W$ and $Z$\ backgrounds by 40\% at a cost of a fifth to a
third of the signal.%

\miscplot{met-jt}{115 211 378 492}%
{Distribution of $\had$}{%
Distribution of $\had$ in the \metjet\ channel for 
150 GeV $\tilde{t}_1 \to$ 80 GeV $\widetilde Z_1$ and backgrounds
\wtn\ and \znn, after applying cut $(vi)$.
Figure elements are as in Fig.~\ref{fig:met-phijj}.
}

Further cuts based on simple kinematical quantities do not seem to be
available. In order to extend the useful discovery reach in this channel we
investigate the possibility of discriminating the $\tau $s in the $%
W\rightarrow \tau \nu $\ background. Since we have already vetoed events in
this channel with an identified lepton, we expect most of these $\tau $s to
decay hadronically, producing jets with low charge multiplicity. For the
purposes of our simulation, we calculate the charge multiplicity of a jet, $%
\mathrm{mult}(j_{{}})$, as the number of long-lived charged hadrons with $E_{%
\mathrm{\top }}^{\mathstrut }>0.5\mathrm{\ GeV}$ lying in the jet cone.
Distributions of the minimum charge multiplicity, $\mathrm{min}_{j}\mathrm{\
mult}(j)$, are shown in Fig.~\ref{fig:met-tau}. Bare $\tau $s will
canonically decay to one or three charged hadrons, but upon analyzing
simulated events we find that a ``$\tau $-veto'' requirement 
\begin{eqnarray*}
\left( viii\right) \quad {\mathrm{min}_{j}\mathrm{\ mult}(j)>4}\quad 
\end{eqnarray*}
gives the best separation of signal and background over the SUSY parameter
space, eliminating almost 60\% of the background along with 30--40\% of the
signal. Note that 70\% of the $t\bar{t}$ background also fails this cut
since $t$ decays with large $\NEG{E}\mathstrut _{\top }^{\text{ }}$ (from an
energetic $\nu $)\ and no visible leptons will usually involve $W\rightarrow
\tau \nu $.%

\miscplot{met-tau}{115 211 378 492}%
{Distribution of $\minmul$}{%
Distribution of $\minmul$ in the \metjet\ channel for 
150 GeV $\tilde{t}_1 \to$ 80 GeV $\widetilde Z_1$ and backgrounds
\wtn\ and \znn, after applying cut $(vii)$.
Figure elements are as in Fig.~\ref{fig:met-phijj}.
}

\conplot{\metjet\ }{
{Cross-section contours, in fb, for the }$\NEG{E}\mathstrut
_{\top }^{\text{ }}${\ + jets\ channel after cut }$\left( viii\right) ${\ as
described in the text. The heavy solid line at 48 fb is the 5$\sigma $
discovery limit for an integrated luminosity of 2 $\mathrm{fb}^{-1}$. The
dashed line is 45 fb of signal, which is 25\% of background. The dot-dashed
line shows the reach at 25 $\mathrm{fb}^{-1}$\ after cut }$\left( ix\right) $%
. The hatched region is
experimentally excluded at the 95\% confidence level.
}
The results of applying cuts $\left( i\right) $--$\left( viii\right) $\ are
displayed as a contour plot in Fig.~\ref{fig:met:con}. This plot shows the
SUSY parameter space $m_{{\tilde{t}_{1}}}$--$m_{{\widetilde{Z}_{1}}}$. The
two diagonal lines indicate the kinematical boundaries of the decay ${\tilde{%
t}_{1}}\rightarrow c{\widetilde{Z}_{1}}$; the upper left hand corner is
excluded by the requirement $m_{{\widetilde{Z}_{1}}}<m_{{\tilde{t}_{1}}}$
that ${\widetilde{Z}_{1}}$ be the LSP, and in the lower right corner the
3-body decay ${\tilde{t}_{1}}\rightarrow bW{\widetilde{Z}_{1}}$ opens up.
The combined LEP\ II/Tevatron Run I excluded area (see Fig.~\ref{fig:expt})
is denoted by the hatched region. The legend in the upper left corner shows
the estimated background cross-sections, in fb, after cut $\left(
viii\right) $. The contour labels are also given in fb. The heavy solid line
at 48 fb represents $N_{s}/\sqrt{N_{b}}=5\sigma $ for the Run II integrated
luminosity of 2 $\mathrm{fb}^{-1}$. The heavy dashed line is at 45 fb = 25\%
of the 180 fb total background. From the plot we see that even with our
conservative criteria, experiments at the Tevatron can discover most cases
with $m_{{\widetilde{Z}_{1}}}\lesssim 70-80\mathrm{\ GeV}$ and $m_{{\tilde{t}%
_{1}}}\lesssim 165$ GeV in this channel.

Note that the contours of Fig.~\ref{fig:met:con} extend into the region
marked ${\tilde{t}_{1}}\rightarrow bW{\widetilde{Z}_{1}}$, where the
three-body decay is kinematically open. Above the ${\tilde{t}_{1}}%
\rightarrow bW{\widetilde{Z}_{1}}$ line the branching fraction $\mathcal{B}({%
\tilde{t}_{1}}\rightarrow c{\widetilde{Z}_{1}})$ is taken to be $100\%$,
while below the line the two decay modes are expected to compete\cite
{porod,porodWohr}.

\subsection{Run II+}

This channel is background limited in that even for higher integrated
luminosity the search is limited by the signal to background ratio, and
without refining the requirements on event topology, such as the $c$-tagging
studied in the next section, little improvement seems available for the high
luminosity scenario of the Run II+ Tevatron. The availability of more
events, though, allows us to make deeper cuts and still preserve an
acceptable $N_{s}/\sqrt{N_{b}}$. By tightening the previous cuts to 
\begin{eqnarray*}
\left( vi^{\prime }\right)  &&{\Delta \phi (j_{1},j_{2})>120^{\circ },} \\
\left( vii^{\prime }\right)  &&{J_{\mathrm{T}}>225\mathrm{\ GeV},} \\
\left( viii^{\prime }\right)  &&{\mathrm{min}_{j}\mathrm{\ mult}(j)>5}
\end{eqnarray*}
we can push the discovery limits at the heavy ${\tilde{t}_{1}}$ end of the
parameter space, up to about 185 GeV. Our projection of the reach at 25 $%
\mathrm{fb}^{-1}$ after these cuts\ is shown on the contour plot of Fig.~\ref
{fig:met:con} as a heavy dot-dashed line. We see that the reach extends
beyond $m_{{\tilde{t}_{1}}}=$ 180 GeV.

\section{$%
\lowercase{c}%
$-tagged\ channel}

In the $c$-tagged\ channel we attempt to identify the charm quark by looking
for a jet with a soft muon due to the decay $c\rightarrow s\bar{\mu}\nu
_{\mu }$. The backgrounds here are similar to those in the previous channel. 
$Z\rightarrow \nu \bar{\nu}$ contributes when accompanied by tagged heavy
flavor production (e.g., $g\rightarrow c\bar{c}$ from initial state QCD
radiation). $W\rightarrow e,\mu $ and $W\rightarrow \tau \rightarrow e,\mu $
are a problem when the lepton mis-tags an unrelated jet. $t\bar{t}$ events
will be a reducible background when a semileptonic $b$-jet is tagged as a $c$%
-jet.

\subsection{Run I}

Our earlier analysis of this channel geared for the Run I\cite{BST}
integrated luminosity suggests cuts as in the Run I $\NEG{E}\mathstrut
_{\top }^{\text{ }}+$ jets channel (see Sect.~\ref{sec:met:runI}), except
that we here require a tagging muon with $p\mathstrut _{\top }^{\text{ }%
}\left( \mu \right) >3$ GeV within $\Delta R=0.4$ of a jet axis, and the
compound cut (4) is changed to (4') $\Delta \phi ({\NEG{E}\mathstrut _{\top
}^{\text{ }},}$nearest jet${)\leq 90}^{\circ }$ or ${p\mathstrut _{\top }^{%
\text{ }}}\left( \text{fast jet}\right) >50$ GeV. We concluded that with the
integrated luminosity available in Run I, the $c$-tagged channel would not
extend the reach beyond the $\NEG{E}\mathstrut _{\top }^{\text{ }}+$ jets
channel because, although it has a better signal to background ratio, the
signal is severely rate limited.

\looseness=2
The CDF Collaboration has analyzed 88 pb$^{-1}$ of Run I data\cite{CDFscz}
looking for $\NEG{E}\mathstrut _{\top }^{\text{ }}+$ $c$-tagged jet with a
different technique than ours, using the SVX detector to identify charmed
hadrons by looking for displaced vertices. They required $\NEG{E}\mathstrut
_{\top }^{\text{ }}>40$ GeV, 2 or 3 (hard)\ jets with $E\mathstrut _{\top }^{%
\text{ }}>15$ GeV within $\left| \eta \right| <2$ (to reject $t\bar{t}$
events, which usually have 4 or more hard jets) but no (soft)\ jets with $%
E\mathstrut _{\top }^{\text{ }}<15$ GeV (to suppress QCD multijet events), $%
45^{\circ }<{\Delta \phi (\NEG{E}\mathstrut _{\top }^{\text{ }},j)<165}%
^{\circ }$, $45^{\circ }<{\Delta \phi (j_{1},j_{2})<165}^{\circ }$, and no
leptons with ${p\mathstrut _{\top }^{\text{ }}>10\mathrm{\ GeV}}$. Then they
look for the SVX charm tag. By this means, they get a 95\% confidence level
excluded region bounded roughly by $m_{{\tilde{t}_{1}}}<115$ GeV and $m_{{%
\widetilde{W}_{1}}}<50$ GeV.

\subsection{Run II}

Initial cuts for the $c$-tagged\ channel are:%
%
%
\begin{eqnarray*}
\left( i\right)  &&\quad \NEG{E}\mathstrut _{\top }^{\text{ }}>50\text{ GeV;}
\\
\left( ii\right)  &&\quad \text{two jets with }p\mathstrut _{\top }^{\text{ }%
}>30\text{ GeV, one with }\left| \eta \right| <1\text{;} \\
\left( iii\right)  &&\quad {\text{for all jets, }\Delta \phi (\NEG%
{E}\mathstrut _{\top }^{\text{ }},j)>30^{\circ },} \\
&&\quad \quad \quad \quad {\text{and if only two jets then }\Delta \phi
(j_{1},j_{2})<150^{\circ };} \\
\left( iv\right)  &&\quad \text{one muon with }p\mathstrut _{\top }^{\text{ }%
}>2\text{ GeV, }\left| \eta \right| <1.7\text{ within }\Delta R=0.4\text{ of
a jet;} \\
\left( v\right)  &&\quad \text{no other visible leptons;} \\
\left( vi\right)  &&\quad \text{no SVX }b\text{ tag.}
\end{eqnarray*}
Cuts $\left( i\right) $\ to $\left( iii\right) $\ are implemented for the
same reasons as in the previous channel. The charm tag lets us relax our jet 
$p\mathstrut _{\top }^{\text{ }}$ requirement. Cut $\left( iv\right) $\
identifies the muon tagging a $c$-jet. After this cut, some background
events will have genuine $c$-jets, others will have $\mu $-tagged $b$
quarks, and yet others will get fake $c$-tags due to miscellaneous muons
accidentally coinciding with unrelated jets. Most $t\bar{t}$\ events in this
sample are of the $b\rightarrow \mu $ sort. Significant numbers of $%
Z\rightarrow \nu \bar{\nu}$\ events are also of this type, since $%
Z\rightarrow \nu \bar{\nu}$\ events which pass the initial cuts are almost
all due to gluon splitting to heavy flavors, and $g\rightarrow b\bar{b}$
occurs roughly as often as $g\rightarrow c\bar{c}$ for these events after
the hard cuts $\left( i\right) $\ and $\left( ii\right) $. We veto extra
isolated leptons and identified $b$-jets with cuts $\left( v\right) $ and $%
\left( vi\right) $.

\cutgroup{cht}
\begin{table} \centering%
\begin{tabular}{|c|c|c|c|}
\hline
background & $\left( vi\right) $ & $\left( vii\right) $ & $\left(
viii\right) $ \\ \hline
$W\rightarrow \tau \nu $ & \multicolumn{1}{|r|}{34.7} & \multicolumn{1}{|r|}{
16.7} & \multicolumn{1}{|r|}{7.1} \\ 
$W\rightarrow \ell \nu $ & \multicolumn{1}{|r|}{61.3} & \multicolumn{1}{|r|}{
9.9} & \multicolumn{1}{|r|}{4.8} \\ 
$Z\rightarrow \nu \bar{\nu}$ & \multicolumn{1}{|r|}{20.5} & 
\multicolumn{1}{|r|}{14.2} & \multicolumn{1}{|r|}{6.6} \\ 
$t\bar{t}$\quad (175) & \multicolumn{1}{|r|}{22.5} & \multicolumn{1}{|r|}{
11.4} & \multicolumn{1}{|r|}{4.4} \\ \hline
total & \multicolumn{1}{|r|}{139.0} & \multicolumn{1}{|r|}{52.2} & 
\multicolumn{1}{|r|}{22.9} \\ \hline
25\% & \multicolumn{1}{|r|}{34.7} & \multicolumn{1}{|r|}{13.0} & 
\multicolumn{1}{|r|}{5.8} \\ 
5$\sigma $ @ 2 fb$^{-1}$ & \multicolumn{1}{|r|}{41.7} & \multicolumn{1}{|r|}{
25.5} & \multicolumn{1}{|r|}{(4.8)} \\ \hline\hline
${\tilde{t}_{1},\widetilde{Z}_{1}}$ & $\left( vi\right) $ & $\left(
vii\right) $ & $\left( viii\right) $ \\ \hline
120,80 & \multicolumn{1}{|r|}{59.9} & \multicolumn{1}{|r|}{48.1} & 
\multicolumn{1}{|r|}{33.5} \\ 
160,90 & \multicolumn{1}{|r|}{36.9} & \multicolumn{1}{|r|}{30.3} & 
\multicolumn{1}{|r|}{19.3} \\ 
180,110 & \multicolumn{1}{|r|}{19.1} & \multicolumn{1}{|r|}{15.9} & 
\multicolumn{1}{|r|}{10.0} \\ \hline
\end{tabular}
\caption[$c$-tagged channel cross-sections]{%
$c$-tagged channel cross-sections, in fb, after cuts as
discussed in the text. The line labelled ``25\%'' shows 25\% of the total
background, and the ``$5\sigma$" line shows the signal level needed to
produce a $5\sigma $ excess at 2 $fb^{-1}$.  The parenthesized $5\sigma $
value in the last column is for an integrated luminosity of 25 $fb^{-1}$.
For the signal cases, $\tilde{t}_{1}$ and $\widetilde{Z}_{1}$ masses are
given in GeV} \label{tab:cht}%
\end{table}%

\def\epsfsize#1#2{\textwidth}
\begin{figure}[tbp!]
\begin{center}
\leavevmode
\makebox[0pt]{\epsffile[45 430 535 900]{Graf/cht.tag.eps}}
\end{center}
\caption[Scatter plots of $p_{\mathrm{T}}^{\mathrm{rel}}$ vs
$\Delta R(\mu,j)$]{\small
Scatter plots of $p_{\mathrm{T}}^{\mathrm{rel}}$ vs $\Delta
R\left( \mu ,j\right) $ for backgrounds and selected signal points in the $c$%
-tagged channel after cut $\left( vi\right) $. \ The diagonal line shows the
cut $\left( vii\right) $ line $\Delta R+p_{\mathrm{T}}^{\mathrm{rel}}/5$ GeV$%
=0.4$.
}
\label{fig:cht:tag}
\end{figure}

After these cuts, we still have the reducible background from non-SVX-tagged 
$b$-jet production as well as fake $c$-tags, in addition to the irreducible
background with real $c$-tags. To reduce the former, we consider the
transverse momentum of the muon with respect to its tagged jet. This is 
\begin{eqnarray*}
p\mathstrut _{\top }^{\text{rel}}\equiv \left| \vec{p}(\mu )\times \hat{p}%
(j)\right| , 
\end{eqnarray*}
where $\hat{p}(j)=\vec{p}(j)/|\vec{p}(j)|$ is the jet direction unit vector.
Since $m_{b}>m_{c}$, the average $\left\langle p\mathstrut _{\top }^{\text{%
rel}}\right\rangle $ will be greater for $b$-jet samples than for $c$-jet
samples. Figure \ref{fig:cht:tag} displays scatter plots of $p\mathstrut
_{\top }^{\text{rel}}$ versus the $\mu $--jet axis separation $\Delta R(\mu
,j)$ for the backgrounds together with a pair of representative signal
samples. Also shown is the line $\Delta R+p_{\mathrm{T}}^{\mathrm{rel}}/5%
\mathrm{\ GeV}=0.4$ which we have selected for our cut. The efficient
separation is obvious from the diagram, and we enhance the $c$-jet purity by
requiring%
%
%
\begin{eqnarray*}
\left( vii\right) \quad {\ \Delta R(\mu ,j)\ +\ p\mathstrut _{\top }^{\text{%
rel}}(\mu |j)/5\mathrm{\ GeV}<0.4.} 
\end{eqnarray*}
As shown in Table \ref{tab:cht}, this cut eliminates 60\% of the background
at a typical cost of 20\% in signal. The large reduction shown in the table
for $W\rightarrow \ell \nu $\ is due to the limit on $p\mathstrut _{\top }^{%
\text{rel}}$, since the high-$p\mathstrut _{\top }^{\text{ }}$ muons in
these events are only coincidentally associated with a jet. We have checked
further cuts, such as limits on $p\mathstrut _{\top }^{\text{ }}(\ell )$,
and found that they have nothing substantial to add after cut $\left(
vii\right) $.

\conplot{\ctagged}{
{Cross-section contours, in fb, for the $c$-tagged\ channel
after cut }$\left( vii\right) ${\ as described in the text. The heavy solid
line at 25.5 fb is the 5$\sigma $ discovery limit for an integrated
luminosity of 2 $\mathrm{fb}^{-1}$. The dashed line is 13 fb of signal,
which is 25\% of background. The dot-dashed line shows the reach at 25 $%
\mathrm{fb}^{-1}$\ after cut }$\left( viii\right)$. The hatched region is
experimentally excluded at the 95\% confidence level.
}

A contour plot summarizing the result of applying cuts $\left( i\right) $-$%
\left( vii\right) $\ is shown in Fig.\ \ref{fig:cht:con}. The total
background remaining at this level is 52.2 fb. The heavy solid line in the
figure gives the 25.5 fb signal level, which corresponds to 5$\sigma $ at an
integrated luminosity of 2 $\mathrm{fb}^{-1}$. This channel affords
substantially more coverage than the $\NEG{E}\mathstrut _{\top }^{\text{ }}$
+ jets channel of the previous section, and much of the parameter space $m_{{%
\tilde{t}_{1}}}\lesssim 170\mathrm{\ GeV}$, $m_{{\widetilde{Z}_{1}}}\lesssim
90\mathrm{\ GeV}$ is accessible to detection.

Some recent analyses\cite{CDFscz,Regina} have looked at tagging charm jets
directly with the SVX. The CDF Collaboration has done this for the Run I
data, yielding the excluded regions shown on Figs.~\ref{fig:met:con} and \ref
{fig:cht:con}. They rely on improved resolution to identify the small impact
parameters of $c$-decay displaced vertices. This analysis has been extended
for the Run II scenario\cite{Regina}. These searches proceed in a
complementary channel to the one we investigate here, since they accept SVX
tags and reject identified leptons (to reduce $W$ and $Z$ backgrounds) while
we veto SVX tags and require a lepton. The CDF estimated reach with these
cuts is similar in extent to ours. It may be worth investigating whether
combining the two $c$-tagging techniques would further extend the Tevatron
reach.

\bigskip\medskip
\subsection{Run II+}
\medskip

\miscplot{cht-dist}{113 127 474 587}%
{Distributions of $\phimc$ and $\njets$}{%
Distributions of $\phimc$ and $\njets$ in the \ctagged\ channel for
 a signal case of 180 GeV $\tilde{t}_1 \to$ 110 GeV $\widetilde Z_1$ and
  backgrounds \ttbar, \znn, \wln\ and \wtn, after cut $(vii)$.
The vertical axes are in fb/bin.
}

Unlike the $\NEG{E}\mathstrut _{\top }^{\text{ }}$ + jets\ channel, the
signal in the $c$-tagged channel is not background limited at 2 $\mathrm{fb}%
^{-1}$, as indicated by the large gap between the solid $5\sigma $ line at
25.5 fb and the dashed $N_{s}/N_{b}=25\%$ line at 13 fb. Thus, more
integrated luminosity can rapidly improve the reach with these same cuts. By
eqn.~(\ref{eq:400}), 7.7 fb$^{-1}$ will bring us to the 13 fb contour. After
this point further cuts are needed for the Run II+ scenario.

To cut the background further, we note that by vetoing events with
non-tagging leptons we have increased the fraction of hadronically decaying $%
t$s, so that most $t\bar{t}$\ events surviving cut $\left( vii\right) $\ are
either $t\rightarrow W\rightarrow \tau $ decays or $\ell \nu $ + 4-jet
events with the lepton faking a charm tag. About 60\% of these events have 5
or more reconstructed jets, as seen in Fig.~\ref{fig:cht-dist}, so limiting
the jet multiplicity helps suppress $t\bar{t}$.

As in the previous channel, the jets in $Z\rightarrow \nu \bar{\nu}$ recoil
against the $Z$ boson, which itself is converted to $\NEG{E}\mathstrut
_{\top }^{\text{ }}$. Thus the angle $\Delta \phi (c,\NEG{E}\mathstrut
_{\top }^{\text{ }})$ between the tagged jet and $\NEG{E}\mathstrut _{\top
}^{\text{ }}$ will be large. For ${\tilde{t}_{1}^{\ast }\tilde{t}_{1}}$
there are two $c$-jets recoiling independently against their two LSPs, and
the distribution of $\Delta \phi (c,\NEG{E}\mathstrut _{\top }^{\text{ }})$
should be broader. This is indeed the case, as shown in Fig.~\ref
{fig:cht-dist}. In $W\rightarrow \ell \nu $ events with $p\mathstrut _{\top
}^{\text{ }}\left( W\right) \ll m_{W}$, the angle $\Delta \phi \left( \ell
,\nu \right) $ will be large and if $\ell $ mistags a jet and $\nu $
supplies $\NEG{E}\mathstrut _{\top }^{\text{ }}$ then $\Delta \phi (c,\NEG%
{E}\mathstrut _{\top }^{\text{ }})$ will be large too. We address these
points by insisting that $\NEG{E}\mathstrut _{\top }^{\text{ }}$ and the $c$%
-jet not be too back-to-back (75\% of the Drell-Yan\ background has $\Delta
\phi (c,\NEG{E}\mathstrut _{\top }^{\text{ }})$ greater than $120^{\circ }$%
). Our Run II+ optimized high luminosity cut is%
%
%
\begin{eqnarray*}
\left( viii\right) \quad \Delta \phi (c,\NEG{E}\mathstrut _{\top }^{\text{ }%
}){<145^{\circ }\text{ and}\ n_{\mathrm{jets}}<5} 
\end{eqnarray*}
which eliminates over half of the background events and fewer than 1/3 of
the signal events. Cross-sections after this cut are shown in Table \ref
{tab:cht}. The dot-dashed discovery line in Fig.~\ref{fig:cht:con} indicates
that ${\tilde{t}_{1}}$s as heavy as 200 GeV may be accessible to the Run II+
Tevatron in this channel, even if the ${\widetilde{Z}_{1}}$ mass is well
beyond 100 GeV.

It is amusing to note, with reference to Fig.~\ref{fig:cht:con}, that if
(1)\ the electroweak baryogenesis story of Section \ref{sec:stop:theo} is
correct and $m_{{\tilde{t}_{1}}}\lesssim 160$ GeV, and (2) the galactic halo
story of Section \ref{sec:stop:theo} is correct and $m_{{\tilde{t}_{1}}}-m_{{%
\widetilde{Z}_{1}}}\gtrsim 20$ GeV, and (3)\ we ``double'' the integrated
luminosity by combining the CDF and D0 experimental results, and (4)\ the
stop decays as ${\tilde{t}_{1}}\rightarrow c{\widetilde{Z}_{1}}$ without the
4-body decay of Section \ref{sec:stop:prod} having a significant branching
fraction, \emph{then} the stop should be discovered at the Run II+ Tevatron!

\chapter{The decay mode $\widetilde{{%
\lowercase{t}%
}}{_{1}}\rightarrow 
\lowercase{b}%
{\widetilde{W}_{1}}$%
\label{chap:sbw}%
}

When $m_{{\tilde{t}_{1}}}>m_{b}+m_{{\widetilde{W}_{1}}}$, the stop decays
via ${\tilde{t}_{1}}\rightarrow b{\widetilde{W}_{1}}$ with essentially 100\%
branching fraction since competing decays occur only at higher order. The ${%
\widetilde{W}_{1}}$ then cascades to a ${\widetilde{Z}_{1}}$ plus a SM
fermion pair. Unlike the ${\tilde{t}_{1}}\rightarrow c{\widetilde{Z}_{1}}$
mode, where the analysis depended only on $m_{{\tilde{t}_{1}}}$ and $m_{{%
\widetilde{Z}_{1}}}$, here we have 4 parameters: the three masses $m_{{%
\tilde{t}_{1}}}$, $m_{{\widetilde{W}_{1}}}$, $m_{{\widetilde{Z}_{1}}}$ and
the leptonic branching fraction of the chargino, $\mathcal{B}({\widetilde{W}%
_{1}}\rightarrow e,\mu \text{ or }\tau )$ (assuming lepton universality in ${%
\widetilde{W}_{1}}$decays). To make the parameter space tractable, we will
adopt the gaugino unification relation $m_{{\widetilde{W}_{1}}}\approx 2m_{{%
\widetilde{Z}_{1}}}$ (eqn.~(\ref{eq:sugra:winozino})) for most of the
discussion, and separately examine the consequences of relaxing this. We
also will take $\mathcal{B}({\widetilde{W}_{1}}\rightarrow \ell )\approx 
\mathcal{B}({W}\rightarrow \ell )$ for the main analysis.

When both ${\widetilde{W}_{1}}$'s decay hadronically, the signal is jets
plus $\NEG{E}\mathstrut _{\top }^{\text{ }}$ from the pair of ${\widetilde{Z}%
_{1}}$'s. This signal, with degraded $\NEG{E}\mathstrut _{\top }^{\text{ }}$
due to the two-step decay of the ${\tilde{t}_{1}}$, is difficult to
discriminate from SM processes. Therefore, we focus on channels where one or
both ${\widetilde{W}_{1}}$'s decay leptonically. In the case of one leptonic
decay we look for a hard isolated lepton together with a $b$-tag along with
the usual $\NEG{E}\mathstrut _{\top }^{\text{ }}$. We refer to this search
as the ``$b$-jet + lepton\ channel''. In the ``dilepton channel'' we look
for $\NEG{E}\mathstrut _{\top }^{\text{ }}$ plus 2 unlike-sign isolated
leptons ($e$ or $\mu $) as the result of both ${\widetilde{W}_{1}}$'s
decaying to leptons.

The dominant background in both channels is SM $t\bar{t}$\ production and
decay followed by the leptonic decay of one or both $W$'s; such processes
have $b$-jets, leptons and substantial $\NEG{E}\mathstrut _{\top }^{\text{ }%
} $ from the associated neutrinos. Indeed, the decays $t\rightarrow
bW\rightarrow b\bar{f}f^{\prime }$ and ${\tilde{t}_{1}}\rightarrow b{%
\widetilde{W}_{1}}\rightarrow b\bar{f}f^{\prime }{\widetilde{Z}_{1}}$ differ
only in the presence of two ${\widetilde{Z}_{1}}$'s in ${\tilde{t}_{1}}%
{}^{\!\ast }{\tilde{t}_{1}}$ events and the presence of on-shell $W$s in $t%
\bar{t}$\ events; event topologies from stop and top production are
identical. The two massive ${\widetilde{Z}_{1}}$s in the final state lead to
a distinctive softening of the spectra of many kinematical quantities in the
signal events. The on-shell $W$ in $t\rightarrow bW$ events implies useful
kinematical constraints on its daughters $\bar{f},f^{\prime }$. This is
especially helpful when only one of the $W$s decays leptonically, in which
case the transverse mass of the lepton and $\NEG{E}\mathstrut _{\top }^{%
\text{ }}$ help to discern the real $W$ in the event. The $W$ kinematics are
also useful in suppressing other backgrounds, such as those due to Drell-Yan 
$W$s, $W$ radiation,\ and $W$ pair production.

\section{$%
\lowercase{b}%
$-jet + lepton\ channel}

In the $b$-jet + lepton\ channel, we look for $\NEG{E}\mathstrut _{\top }^{%
\text{ }}$, hard jets with a tagged $b$, and an isolated $e$ or $\mu $. In
this channel we contend with two main SM backgrounds. The first is $t\bar{t}$%
\ production, as mentioned above. The second is due to events with a
leptonically decaying weak boson $W\rightarrow e,\mu ,\tau $ associated with
QCD radiation. This reaction has a huge cross-section, and readily delivers
large $\NEG{E}\mathstrut _{\top }^{\text{ }}$ from the decay neutrino, along
with a hard isolated lepton. Much of this background is suppressed by
requiring a tagged $b$-jet in the event. The $W$ jets that pass this
criterion generally have $b$s from gluon splitting $g\rightarrow b\bar{b}$.
A minor background comes from the process $Z\rightarrow \tau ^{+}\tau ^{-}$\
when one of the $\tau $s decays leptonically and a real or fake $b$ is
tagged. We ignore mistagged $b$-jet backgrounds.

\subsection{Run I}

In our earlier pre-LEP analysis\cite{BST} of the Run I situation for this
channel, we suggested the following initial cuts: (1) $\NEG{E}\mathstrut
_{\top }^{\text{ }}>25$ GeV; (2)\ at least 2 jets with ${p_{\mathrm{T}%
}^{\mathstrut }>15\mathrm{\ GeV}}$, one of which lies in the central region $%
{|\eta |<2}$; (3)\ an isolated electron or muon with ${p_{\mathrm{T}%
}^{\mathstrut }}\left( e\right) >10$ GeV and ${p_{\mathrm{T}}^{\mathstrut }}%
\left( \mu \right) >5$ GeV; and (4)\ an SVX $b$-tag (with an efficiency
estimated at 40\% (in the barrel)\ $\times $ 30\% (identified displaced
vertex)\ = 13\%)\footnote{%
In the meantime, SVX technology has improved considerably. Resolution has
gone up and coverage has extended to $\left| \eta \right| <2$. For Run II,
both D0 and CDF will be equipped with SVX detectors.}. The purpose of these
cuts is the same as the analogous cuts in our Run II analysis presented
below. After these, we also recommended (5)\ no more than 4 jets (with ${p_{%
\mathrm{T}}^{\mathstrut }>15\mathrm{\ GeV}}$); and (6)\ the transverse mass
cut $m\mathstrut _{\top }^{\text{ }}(\ell ,\NEG{E}\mathstrut _{\top }^{\text{
}})<45$ GeV (see below for this quantity). We found that in this channel the
signal dwindles as $m_{{\tilde{t}_{1}}}\rightarrow m_{{\widetilde{W}_{1}}}$,
where the $b$ becomes too soft to tag. We determined that the region $m_{{%
\tilde{t}_{1}}}\lesssim 90$ GeV, $m_{{\widetilde{W}_{1}}}\lesssim 60$ GeV
would be accessible to this treatment with the Run I integrated luminosity
of 100 pb$^{-1}$.

The CDF Collaboration examined this channel for $88\pm 4$ pb$^{-1}$ of their
Run I data\cite{CDFsbw}. They used the cuts (1)\ $\NEG{E}\mathstrut _{\top
}^{\text{ }}>25$ GeV; (2)\ at least 2 jets with ${p_{\mathrm{T}}^{\mathstrut
}>}$ 12 GeV for the hardest jet and ${p_{\mathrm{T}}^{\mathstrut }>}$ 8 GeV
for the second jet; (3)\ an isolated electron or muon with ${p_{\mathrm{T}%
}^{\mathstrut }>10}$ GeV; and an SVX $b$-tag. They further required that (5)
any dilepton have invariant mass $m\left( \ell ^{+}\ell ^{-}\right) <60$
GeV; and (6)\ the azimuthal angles between $\NEG{E}\mathstrut _{\top }^{%
\text{ }}$ and each of the two hardest jets satisfy $\Delta \phi \left( \NEG%
{E}\mathstrut _{\top }^{\text{ }},\text{jet}\right) >0.5$ rad. They found no
evidence for ${\tilde{t}_{1}^{\ast }\tilde{t}_{1}}$ production, but were not
able to improve the existing (LEP\ II) stop mass limits.

\subsection{Run II}

For the higher luminosity run, our initial cuts for this channel are 
\begin{eqnarray*}
\left( i\right) &&\quad \NEG{E}\mathstrut _{\top }^{\text{ }}{>25\mathrm{\
GeV};} \\
\left( ii\right) &&\quad {\text{exactly one isolated $e$ or $\mu $ with}\ p_{%
\mathrm{T}}^{\mathstrut }>8\mathrm{\ GeV}\text{ and }|\eta |<2;} \\
\left( iii\right) &&\quad \text{a}{\text{t least two jets with }p_{\mathrm{T}%
}^{\mathstrut }>15\mathrm{\ GeV}\text{ and }|\eta |<2;} \\
\left( iv\right) &&\quad {\text{at least one SVX $b$-tagged jet with }|\eta }%
_{B}{|<1.4\text{.}}\quad
\end{eqnarray*}
Cuts $\left( i\right) $ to $\left( iii\right) $ are canonical for this
channel. The $b$-tag, cut $\left( iv\right) $, is critical to reduce the $W$
background. The $b$-tagging efficiency for signal events ranges from one
third to two thirds, while less than a percent of the $W$ events get tagged.
The ${|\eta }_{B}{|}$ distribution for $W$ events with $b$s due to $%
g\rightarrow b\bar{b}$ is broader than that for signal events, as shown in
Fig.~\ref{fig:lbb}, hence the cutoff at ${|\eta }_{B}{|<1.4}$. The
soft-lepton $b$-tagging used in SM $t\bar{t}$ studies is avoided here to
avoid the large background from events like $sg\rightarrow Wc\rightarrow
\ell \nu c$ or $q\bar{q}\rightarrow Wg\rightarrow \ell \nu dc\bar{c}$ where $%
\ell $ is isolated, $\nu $ provides $\NEG{E}\mathstrut _{\top }^{\text{ }}$,
the $c$ is tagged and $g$ provides the second jet. (In the $t\bar{t}$
studies, one makes hard cuts to reject these $W$ events; as we shall see
below such hard cuts would remove too much of our stop signal.)

\miscplot{lbb}{94 127 390 519}%
{Distribution of $\left|\eta_B\right|$}{%
Distribution of $\left|\eta_B\right|$ in the \lplusb\ channel
for the initial cut sample.
The signal case is a 160 GeV 
${\tilde{t}_{1}}$ decaying to a 120 GeV 
${\widetilde{W}_{1}}$ which in turn decays to a 60 GeV
${\widetilde{Z}_{1}}$.
The vertical axes are in fb/bin.
}

\cutgroup{lb}
\begin{table} \centering%
\begin{tabular}{|c|l|l|l|l|r|l|}
\hline
background & $\left( iv\right) $ & $\left( v\right) $ & $\left( vi\right) $
& $\left( vii\right) $ & $(viii)$ & $\left( ix\right) $ \\ \hline
$t\bar{t}$\quad (175) & \multicolumn{1}{|r|}{916} & \multicolumn{1}{|r|}{247}
& \multicolumn{1}{|r|}{138} & \multicolumn{1}{|r|}{138} & 80 & 
\multicolumn{1}{|r|}{15.7} \\ 
$W\rightarrow e,\mu ,\tau $ & \multicolumn{1}{|r|}{387} & 
\multicolumn{1}{|r|}{96} & \multicolumn{1}{|r|}{88} & \multicolumn{1}{|r|}{51
} & 48 & \multicolumn{1}{|r|}{5.2} \\ \hline
total & \multicolumn{1}{|r|}{1303} & \multicolumn{1}{|r|}{343} & 
\multicolumn{1}{|r|}{226} & \multicolumn{1}{|r|}{189} & 128 & 
\multicolumn{1}{|r|}{20.9} \\ \hline
25\% & \multicolumn{1}{|r|}{326} & \multicolumn{1}{|r|}{86} & 
\multicolumn{1}{|r|}{56} & \multicolumn{1}{|r|}{47} & 32 & 
\multicolumn{1}{|r|}{5.2} \\ 
5$\sigma $ @ 2 fb$^{-1}$ & \multicolumn{1}{|r|}{128} & \multicolumn{1}{|r|}{
66} & \multicolumn{1}{|r|}{53} & \multicolumn{1}{|r|}{49} & 40 & 
\multicolumn{1}{|r|}{(4.6)} \\ \hline\hline
${\tilde{t}_{1},}$ ${\widetilde{W}_{1}}$ & \multicolumn{1}{|c|}{$\left(
iv\right) $} & \multicolumn{1}{|c|}{$\left( v\right) $} & 
\multicolumn{1}{|c|}{$\left( vi\right) $} & \multicolumn{1}{|c|}{$\left(
vii\right) $} & $(viii)$ & \multicolumn{1}{|c|}{$\left( ix\right) $} \\ 
\hline
140, 120 & \multicolumn{1}{|r|}{134} & \multicolumn{1}{|r|}{72} & 
\multicolumn{1}{|r|}{66} & \multicolumn{1}{|r|}{55} & 50 & 
\multicolumn{1}{|r|}{6.4} \\ 
160, 120 & \multicolumn{1}{|r|}{106} & \multicolumn{1}{|r|}{57} & 
\multicolumn{1}{|r|}{50} & \multicolumn{1}{|r|}{47} & 40 & 
\multicolumn{1}{|r|}{7.4} \\ 
180, 110 & \multicolumn{1}{|r|}{65} & \multicolumn{1}{|r|}{36} & 
\multicolumn{1}{|r|}{29} & \multicolumn{1}{|r|}{29} & 23 & 
\multicolumn{1}{|r|}{5.1} \\ \hline
\end{tabular}
\caption[$b$-jet + lepton channel cross-sections]{%
$b$-jet + lepton channel cross-sections, in fb, after cuts as
discussed in the text. Signal levels required for detection are given in
the ``25\%" and ``5$\sigma$" rows.  The parenthesized $5\sigma$ value
in the last column is for an integrated luminosity of 25 $\text{fb}^{-1}$.
For the signal cases, $\tilde{t}_{1}$ and $\widetilde{W}_{1}$ masses are
given in GeV, and $m_{\widetilde{Z}_1}=m_{\widetilde{W}_1}/2$.
Cut $(ix)$ is made after cut $(viii')$.}
\label{tab:lb}%
\end{table}%

We expect that for $t\bar{t}$ and $W\rightarrow e,\mu $ backgrounds, where
missing transverse energy is due to a single neutrino and $\NEG{E}\mathstrut
_{\top }^{\text{ }}\sim p\mathstrut _{\top }^{\text{ }}\left( \nu \right) $,
the transverse mass $m\mathstrut _{\top }^{\text{ }}(\ell ,\NEG{E}\mathstrut
_{\top }^{\text{ }})$ of the lepton and $\NEG{E}\mathstrut _{\top }^{\text{ }%
}$ should show a strong Jacobian peak near $m_{W}$, while the signal
distribution should be a broad bump. Figure \ref{fig:lbm} shows that this is
indeed the case. The transverse mass is the invariant mass of the $\NEG%
{E}\mathstrut _{\top }^{\text{ }}$ vector and the projection of the lepton's
momentum in the transverse plane, 
\begin{eqnarray*}
m\mathstrut _{\top }^{\text{ }}(\ell ,\NEG{E}\mathstrut _{\top }^{\text{ }%
})^{2}=2p\mathstrut _{\top }^{\text{ }}\left( \ell \right) \NEG{E}\mathstrut
_{\top }^{\text{ }}\left( 1-\cos \Delta \varphi \right) .
\end{eqnarray*}
We can get even more information from this kinematical data by forming a
``semi-transverse'' mass\ which is the invariant mass of $\NEG{E}\mathstrut
_{\top }^{\text{ }}$ and the lepton's full 4-momentum: 
\begin{eqnarray}
m\mathstrut _{\top +}^{\text{ }}(\ell ,\NEG{E}\mathstrut _{\top }^{\text{ }%
})^{2}=2p\mathstrut _{\top }^{\text{ }}\left( \ell \right) \NEG{E}\mathstrut
_{\top }^{\text{ }}\left( 1/\sin \vartheta _{\ell }-\cos \Delta \varphi
\right) 
\end{eqnarray}
where $\vartheta _{\ell }=2\arctan \exp \left( -\eta _{\ell }\right) $ is
the lepton's polar angle. The $m\mathstrut _{\top +}^{\text{ }}$
distribution is also plotted in Fig.~\ref{fig:lbm}, where one sees that for
the $W$-containing backgrounds it has a broader high-end tail than $%
m\mathstrut _{\top }^{\text{ }}$, and that the distribution vanishes in the
limit $m\mathstrut _{\top +}^{\text{ }}\rightarrow 0$, unlike $m\mathstrut
_{\top }^{\text{ }}$ which remains finite there. Both of these behaviors
make $m\mathstrut _{\top +}^{\text{ }}$ a better discriminator than $%
m\mathstrut _{\top }^{\text{ }}$, as the signal distributions for the two
quantities are largely the same. With a cut of 
\begin{eqnarray*}
\left( v\right) \quad m\mathstrut _{\top +}^{\text{ }}(\ell ,\NEG%
{E}\mathstrut _{\top }^{\text{ }}){<60\mathrm{\ GeV}}
\end{eqnarray*}
{\ 3/4 of the }$t\bar{t}$ and $W\rightarrow \ell \nu $ {background is
removed with less than half of the signal lost. (In comparison, an }$%
m\mathstrut _{\top }^{\text{ }}$ cut removes less than 2/3 of the background 
$W$s at the same signal loss.) The results of this cut are listed in Table 
\ref{tab:lb}.%

\miscplot{lbm}{94 127 477 519}%
{Distributions of $\mtlm$ and $\mtlmp$}{%
Distributions of $\mtlm$ and $\mtlmp$ in the \lplusb\ channel,
after cut $(iv)$.
The signal case is a 160 GeV 
${\tilde{t}_{1}}$ decaying to a 120 GeV 
${\widetilde{W}_{1}}$ which in turn decays to a 60 GeV
${\widetilde{Z}_{1}}$.
The vertical axes are in fb/bin.
The dashed lines show cut $(v)$.
}

The $t\bar{t}$\ background still dominates the signal. To reduce it further,
we note that the lepton from $W$ decay is typically more energetic than that
from ${\widetilde{W}_{1}}$ decay. In the decaying $W$ or ${\widetilde{W}_{1}}
$ frame, one can crudely compare the available lepton energy, $m_{W}/2$ for $%
W\rightarrow \ell \nu $, with $(m_{{\widetilde{W}_{1}}}-m_{{\widetilde{Z}_{1}%
}})/2\sim m_{{\widetilde{W}_{1}}}/4$ for ${\widetilde{W}_{1}}\rightarrow
\ell \nu {\widetilde{Z}_{1}}$ and use the fact that generally $m_{{%
\widetilde{W}_{1}}}<2m_{W}$ for the models accessible in this channel. We
have also found that the third jet tends to be harder in $t\bar{t}$\ events
than in our signal events. The reason is similar to that given for the
lepton since, in the absence of hard QCD radiation, this third jet mainly
comes from the hadronically decaying $W$ or ${\widetilde{W}_{1}}$. We
capture this feature by defining the quantity 
\begin{eqnarray}
W\equiv \sqrt{p\mathstrut _{\top }^{\text{ }}(\ell )^{2}+p\mathstrut _{\top
}^{\text{ }}(j_{3})^{2}} 
\end{eqnarray}
where $p\mathstrut _{\top }^{\text{ }}(j_{3})$ is taken as 0 for those
events with only two jets. Imposing the cut 
\begin{eqnarray*}
\left( vi\right) \quad W<\text{60 GeV} 
\end{eqnarray*}
reduces the dominant background from $t\bar{t}$ by almost half and the
signal by just 10--20\%. $W$ distributions are displayed in Fig.~\ref
{fig:lbw}, and the results of the cut are tabulated in Table \ref{tab:lb}.%

\miscplot{lbw}{93 127 390 519}%
{Distribution of $W$}{%
Distribution of $W$ in the \lplusb\ channel, after cut $(v)$.
The signal case is a 160 GeV 
${\tilde{t}_{1}}$ decaying to a 120 GeV 
${\widetilde{W}_{1}}$ which in turn decays to a 60 GeV
${\widetilde{Z}_{1}}$.
The vertical axes are in fb/bin.
The dashed lines at $W=60$ GeV indicate cut $(vi)$.
}

We turn our attention now to the $W$ events, which are distinguished in
several ways from the signal (and $t\bar{t}$) events. First, their jets come
from QCD radiation so they have a lower jet multiplicity. Also, they have a
low efficiency for a second $b$ tag, a fact used in the Run II+ section
below. Finally, the $W$ and $Z$\ events are generally softer than the signal
events, which is reflected in the distribution of the quantity 
\begin{eqnarray}
H_{T+}\equiv \NEG{E}\mathstrut _{\top }^{\text{ }}+p\mathstrut _{\top }^{%
\text{ }}(\ell )+\sum_{j}p\mathstrut _{\top }^{\text{ }}(j),
\end{eqnarray}
which is well-known from $t\bar{t}$\ studies to provide a good separation of 
$t\bar{t}$\ from the $W\rightarrow e,\mu ,\tau $\ background. Figure \ref
{fig:lbh} shows these distributions. The ${\tilde{t}_{1}}$ distribution
unfortunately falls in between the two backgrounds, a predicament which is
quite general across the parameter space we have investigated for this
channel. Still, the cut 
\begin{eqnarray*}
\left( vii\right) \quad {H_{T+}>}140\text{ GeV}
\end{eqnarray*}
eliminates 40\% of the $W$ background with only a few percent loss of
signal, as shown in Table \ref{tab:lb}.%

\miscplot{lbh}{93 127 390 519}%
{Distribution of $\htp$}{%
Distribution of $\htp$ in the \lplusb\ channel after cut $(vi)$.
The signal case is a 160 GeV 
${\tilde{t}_{1}}$ decaying to a 120 GeV 
${\widetilde{W}_{1}}$ which in turn decays to a 60 GeV
${\widetilde{Z}_{1}}$.
The vertical axes are in fb/bin.
The dashed line at $\htp=140$ GeV shows cut $(vii)$.
}

So far, we have not used the fact that elaborate techniques have been
developed for reconstructing $t\bar{t}$\ events. The single lepton plus $b$%
-tag channel is a primary source of $t\bar{t}$\ events for Standard Model
top mass reconstruction studies and our experimental colleagues will
certainly subject events in this channel to a thorough characterization
using these techniques. Mass reconstructions begin by assuming that the
missing $E_{T}$ represents the momentum of the $\nu $ from a leptonic $%
W\rightarrow \ell \nu $ decay (the other $W$ is assumed to have decayed
hadronically). By forcing $\nu $ and $\ell $ to reconstruct a real $W$, one
gets two solutions for the longitudinal momentum of the $\nu $. Together
with combinatorics from assigning jets to the (possibly tagged) $b$s, the
hadronic $W$ daughters and miscellaneous QCD radiation, one gets several
fitted solutions. For top mass reconstruction studies, these solutions are
variously all used, used with weights, or judged to select a best solution.
Although straightforward in principle, such analyses are quite complicated
in practice. Results are sensitive functions of jet energy corrections, and
thus depend strongly on the details of the detector simulation.%

\miscplot{lbt}{93 127 390 519}%
{Distribution of $\topx$}{%
Distribution of $\topx$ in the \lplusb\ channel after cut $(vii)$.
The signal case is a 160 GeV 
${\tilde{t}_{1}}$ decaying to a 120 GeV 
${\widetilde{W}_{1}}$ which in turn decays to a 60 GeV
${\widetilde{Z}_{1}}$.
The vertical axes are in fb/bin.
Events that are not reconstructible (for which $\topx=0$) 
are not shown in the plots; the plotted events constitute
82\% of the \ttbar\ sample, 15\% of the \wln\ background, and
46\% of the signal sample.
}

To estimate the utility of using $t\bar{t}$\ characterization to suppress
these events as a background, we have performed toy reconstructions on our
Monte Carlo samples. We use the procedure outlined above, without applying
jet energy corrections. For each event, we judge a best solution to be that
which minimizes the quantity 
\begin{eqnarray}
(m_{t}^{\mathrm{lep}}-m_{t}^{\mathrm{had}})^{2}+(m_{W}^{\mathstrut
}(jj)-m_{W}^{\mathstrut })^{2}, 
\end{eqnarray}
where $m_{t}^{\mathrm{lep}}$ is the reconstructed mass of the semi-leptonic
top, $m_{t}^{\mathrm{had}}$ that of the hadronic top, and $m_{W}^{\mathstrut
}(jj)$ the invariant mass of the two jets assigned to the hadronic $W$. From
this best solution we form the quantity 
\begin{eqnarray}
\widehat{m}_{t}\equiv (m_{t}^{\mathrm{lep}}+m_{t}^{\mathrm{had}})/2 
\end{eqnarray}
which is our (crude) representation of the fitted top mass. Reconstruction
is not possible for events with less than four jets, and for those events we
define $\widehat{m}_{t}=0$. Events with more than four jets are
reconstructed with all possible jet assignments to find the best fit. The
resulting $\widehat{m}_{t}$ distributions are shown in Fig.~\ref{fig:lbt}.
The crudity of our reconstruction is manifested in the $t\bar{t}$ plot,
which is asymmetric and peaks noticeably below $m_{t}$, largely due to our
lack of jet energy correction. Even at this level of sophistication though
we can reject almost half of the $t\bar{t}$ contamination with a penalty of
less than 15\% of signal by cutting 
\begin{eqnarray*}
\left( viii\right) \quad \widehat{m}_{t}<150GeV. 
\end{eqnarray*}

\conplot{\lplusb}{
{Cross-section contours, in fb, for the $b$-jet + lepton\
channel after cut }$\left( viii\right) ${\ as described in the text. The
heavy solid line at 40 fb is the 5$\sigma $ discovery limit for an
integrated luminosity of 2 $\mathrm{fb}^{-1}$. The dashed line is 32 fb of
signal, which is 25\% of background. The dot-dashed line shows the reach at
25 $\mathrm{fb}^{-1}$\ after cut }$\left( ix\right) ${. We have set $m_{{%
\widetilde{Z}_{1}}}=m_{{\widetilde{W}_{1}}}/2$ for the signal cases.}
}

Figure \ref{fig:lb:con} summarizes the discovery potential of applying cuts $%
\left( i\right) $--$\left( viii\right) $\ in the $b$-jet + lepton\ channel.
The parameter space of this plot is $m_{{\tilde{t}_{1}}}$--$m_{{\widetilde{W}%
_{1}}}$, where we have taken $m_{{\widetilde{Z}_{1}}}=m_{{\widetilde{W}_{1}}%
}/2$ and $\mathcal{B}({\widetilde{W}_{1}}\rightarrow \ell )=\mathcal{B}%
(W\rightarrow \ell )$. The contours show signal cross-sections in fb, on a
total background of 128 fb. The solid line at 40 fb indicates the $5\sigma $
reach for 2 $\mathrm{fb}^{-1}$ of integrated luminosity, which is an
expected 80 SUSY events on top of 256 SM events. The dashed line corresponds
to a quarter of the background cross section. Stops with mass up to $\sim 160%
\mathrm{\ GeV}$ should be discoverable in this channel for charginos lighter
than 120 GeV. Notice from Table \ref{tab:lb} that for lighter stops cuts $%
\left( i\right) $ - $\left( vi\right) $ may give a better significance of
the signal.

\subsubsection{Relaxing the assumption $m_{{\widetilde{Z}_{1}}}=m_{{%
\widetilde{W}_{1}}}/2$}

If we consider models more general than mSUGRA and its cogeners, which have
strict gaugino unification at the GUT scale, then the relationship $m_{{%
\widetilde{Z}_{1}}}=m_{{\widetilde{W}_{1}}}/2$ of eqn.~\ref
{eq:sugra:winozino} may fail to hold\footnote{%
Indeed, this relation is only approximately true even in the mSUGRA
framework because of gaugino-higgsino mixing.}. We study this situation by
performing our analysis in the $m_{{\widetilde{W}_{1}}}-m_{{\widetilde{Z}_{1}%
}}$ plane for the case of a 160 GeV stop. Our results are presented in the
contour map of Fig.~\ref{fig:lb:czw}. The contours are signal cross-sections
in fb after cut $\left( viii\right) $ and we show the 32 fb and 40 fb reach
lines, as in Fig.~\ref{fig:lb:con}. The straight dot-dashed line traces $m_{{%
\widetilde{Z}_{1}}}=m_{{\widetilde{W}_{1}}}/2$, and indicates the slice of
this plot corresponding to the $m_{{\tilde{t}_{1}}}=160$ GeV slice of Fig.~%
\ref{fig:lb:con}. The diagonal line cutting off the lower right corner is
where ${\widetilde{Z}_{1}}$ fails to be the LSP, contrary to our hypothesis.

\czwplot{\lplusb}{
{Cross-section contours, in fb, for the $b$-jet + lepton\
channel in the $m_{{\widetilde{W}_{1}}}$--$m_{{\widetilde{Z}_{1}}}$\ plane
after cut }$\left( viii\right) ${\ as described in the text. The top squark
mass is 160 GeV. Annotations are as in Fig.~\ref{fig:lb:con}.}
}

The straight solid line marked $m_{{\widetilde{W}_{1}}}=m_{{\widetilde{Z}_{1}%
}}+M_{W}$ indicates where the mediating $W^{\ast }$ in the chargino decay
(see Fig.~\ref{fig:winodecay}a and attendant discussion) goes on-shell. The
signal efficiency falls in this region as our $W$-rejection cuts take hold.
Also, when ${\widetilde{Z}_{1}}$ is too light, the fermions from the ${%
\widetilde{W}_{1}}$ decay are energetic enough that the cuts $\left(
v\right) $ and $\left( vi\right) $ designed to suppress $t\bar{t}$\
background erode the signal as well. In the main part of the plot, the
efficiency is fairly independent of the ${\widetilde{Z}_{1}}$ mass, but it
falls again in the limit $m_{{\widetilde{Z}_{1}}}\rightarrow m_{{\widetilde{W%
}_{1}}}$, as the lepton becomes undetectable. In general, the ${\tilde{t}_{1}%
}$ detectability declines outside the region $m_{{\widetilde{W}_{1}}%
}/2\lesssim m_{{\widetilde{Z}_{1}}}\lesssim 2m_{{\widetilde{W}_{1}}}/3$.

\subsubsection{Relaxing the assumption $\mathcal{B}({\widetilde{W}_{1}}%
\rightarrow \ell )\approx \mathcal{B}({W}\rightarrow \ell )$}

We can also ask what happens if the chargino's leptonic branching fraction
is not $W$-like. As discussed in Section \ref{sec:stop:prod}, this is due to 
$\tilde{f}^{\ast }$-mediated decays, which for models with GUT scale
universal scalar masses will usually favor slepton mediation (over squark
mediation)\ due to general features of the RGE. This will increase lepton
production and for moderate values of the enhancement 
\begin{eqnarray}
r=\frac{\mathcal{B}({\widetilde{W}_{1}}\rightarrow \ell )}{\mathcal{B}({W}%
\rightarrow \ell )}%
\label{eq:r}%
\end{eqnarray}
the signal cross-sections should go up by the factor $r$ (without any
increase in backgrounds)\footnote{%
An admixture of the sfermion-mediated diagram could also change the
production kinematics. For instance, if $m_{{\widetilde{W}_{1}}}\approx m_{%
\tilde{\nu}_{\ell }}\ll m_{\tilde{\ell}}$, then the dominant $\tilde{\nu}%
_{\ell }$-mediated decay would produce very soft leptons.}. Of course, in
the limit that ${\widetilde{W}_{1}}$ decays \emph{only} to leptons the
signal will be lost, since we veto the second lepton in this channel. The
single-lepton fraction is maximized when $\mathcal{B}({\widetilde{W}_{1}}%
\rightarrow \ell )=0.5$, or $r\approx 2$. If $r$ were to actually take this
value, then the discovery reach at 2 fb$^{-1}$ would extend to the 20 fb
line of Fig.~\ref{fig:lb:con}, finding stops as heavy as 185 GeV and
charginos as heavy as 140 GeV.

\subsection{Run II+}

\looseness=2
The $N_{S}/N_{B}=25\%$ line at 32 fb in Fig. \ref{fig:lb:con} is well beyond
the $N_{S}/\sqrt{N_{B}}=5\sigma $ line at 40 fb, which indicates that we can
do more with cuts $(i)-(viii)$ if more integrated luminosity is available.
The $5\sigma $ reach would extend out to the 32 fb line with 3.2 fb$^{-1}$
of data (see eqn.~(\ref{eq:400})), revealing stops almost as heavy as the
top quark. For even higher integrated luminosities, we can reduce the $t\bar{%
t}$ background by making a deeper cut on the reconstructed top mass%
\begin{eqnarray*}
\left( viii^{\prime }\right) \quad {\widehat{m}_{t}<125\mathrm{\ GeV.}} 
\end{eqnarray*}
Then we can attack the $W$ background by insisting on a second $b$-tag. In
general, the $W$ event's tagged $b$ jets are less distinguishable since they
are due to gluon splitting ($g\rightarrow b\bar{b}$ frequently forms only
one jet). This leads to a low efficiency for double-tagging; while the
signal and $t\bar{t}$\ events have about a 25\% probability of having a
second $b$ tag, only about 8\% of the $W$ and $Z$\ events get this second
tag. We require a second tag in the Run II+ scenario:%
\begin{eqnarray*}
\left( ix\right) \quad \text{{at least 2 }}b\text{-tags} 
\end{eqnarray*}
At 25 $\mathrm{fb}^{-1}$ of data these cuts give the dot-dashed line in Fig.~%
\ref{fig:lb:con}, where the reach in $m_{{\tilde{t}_{1}}}$ now extends well
beyond $m_{t}$. The reach still cuts off at $m_{{\widetilde{W}_{1}}}\approx
125$ GeV though, presumably because the $b$-jets have to be hard enough in
order to be tagged.

\section{Dilepton channel}

In the dilepton channel we look for two opposite sign hard isolated leptons,
presumed to come from the leptonic decays of both ${\widetilde{W}_{1}}$s in
the ${\tilde{t}_{1}}{}^{\!\ast }{\tilde{t}_{1}}$ event. The major
backgrounds here are $t\bar{t}$, as usual, along with $W$ pair\ production
and the processes $Z\rightarrow \ell ^{+}\ell ^{-}$\ ($\ell =e,\mu $) and $%
Z\rightarrow \tau ^{+}\tau ^{-}$. (Note that $Z\rightarrow \tau ^{+}\tau
^{-} $ was left out of \ the study \cite{BST}.) As in the channels
previously investigated here, the $t\bar{t}$ background is controlled by
rejecting events which are too ``hard.'' The $W$ pair\ process has a modest
production cross-section, but in the absence of $b$-tagging its event
topology is quite similar to that of ${\tilde{t}_{1}}{}^{\!\ast }{\tilde{t}%
_{1}}$ and this background proves the most difficult to remove. Drell-Yan\ $%
Z\rightarrow \ell ^{+}\ell ^{-}$\ has a large production cross-section, but
there is no source of $\NEG{E}\mathstrut _{\top }^{\text{ }}$ in the basic
reaction and those events which remain are effectively controlled by
reconstructing the $Z $ boson from the same-flavor dilepton invariant mass.
Finally, to reduce the background from $Z\rightarrow \tau ^{+}\tau ^{-}$\
with both $\tau $s decaying leptonically we reconstruct the $\tau ^{+}\tau
^{-}$ invariant mass, as explained below.

\subsection{Run I}

For the dilepton channel, our Run I analysis\cite{BST} proposed the
following cuts:\ (1)\ $\NEG{E}\mathstrut _{\top }^{\text{ }}{>25}\mathrm{\
GeV;}$(2)\ at least one jet with ${p\mathstrut _{\top }^{\text{ }}>15\mathrm{%
\ GeV}\text{ and }|\eta |<2;}$ (3) an opposite sign\ ${\ell ^{+}\ell }%
^{\prime }{^{-}}$ pair with${\ }p\mathstrut _{\top }^{\text{ }}\left( {\ell }%
\right) >8(5)$ GeV for $\ell =e(\mu )$; and (4) an azimuthal dilepton angle
satisfying ${20^{\circ }<\Delta \phi (\ell ^{+},\ell ^{-})<160^{\circ }}$.
We also defined a kinematical quantity called bigness, $B$ (see eqn.~(\ref
{eq:bigness})), and imposed the cut (5)\ $B<100$ GeV. Our analysis, which
ignored the important background $Z\rightarrow \tau ^{+}\tau ^{-}$, found a
reach to $m_{{\tilde{t}_{1}}}\lesssim 110$ GeV, $m_{{\widetilde{W}_{1}}%
}\lesssim 90$ GeV. We noted that in distinction to the $b$-jet + lepton
channel the signal here was not strongly attenuated in the limit $m_{{\tilde{%
t}_{1}}}\rightarrow m_{{\widetilde{W}_{1}}}$, since a $b$-tag was not
required.

This study was undertaken by the D0 Collaboration on $75\pm 8$ pb$^{-1}$ of
their Run I data, looking specifically in the dielectron channel\cite{D0sbw}%
, in which the signal is four times smaller than if both electrons and muons
are counted. They used the RGSEARCH cut optimization method and imposed (1)\ 
$\NEG{E}\mathstrut _{\top }^{\text{ }}{>22\mathrm{\ GeV};}$ (2)\ $%
E\mathstrut _{\top }^{\text{ }}\left( {j}_{1}\right) >30$ GeV; (3)\ ${%
p\mathstrut _{\top }^{\text{ }}}\left( e_{1}\right) >16$ GeV and ${%
p\mathstrut _{\top }^{\text{ }}}\left( e_{2}\right) >8$ GeV; and (4)\ $%
m(e^{+}e^{-})<60$ GeV. They also applied our bigness cut as (5)\ $B<90$ GeV.
Two events remained in the data set after these cuts, and the experimenters
found too much $Z\rightarrow \tau ^{+}\tau ^{-}$ background to set a
meaningful limit. Below, we outline a technique which may be used to control
this background.

\subsection{Run II}

The initial cuts for the dilepton channel are 
\begin{eqnarray*}
\left( i\right) &&\quad \NEG{E}\mathstrut _{\top }^{\text{ }}{>25\mathrm{\
GeV};} \\
\left( ii\right) &&\quad {\text{at least one jet with}\ p\mathstrut _{\top
}^{\text{ }}>15\mathrm{\ GeV}\text{ and }|\eta |<2;} \\
\left( iii\right) &&\quad {\text{an opposite-sign }\ell ^{+}\ell ^{-}}\text{
pair with}{\ }p\mathstrut _{\top }^{\text{ }}\left( {\ell }\right) >10\text{
GeV and }{|\eta |<2;} \\
\left( iv\right) &&\quad {20^{\circ }<\Delta \phi (\ell ^{+},\ell
^{-})<160^{\circ };} \\
\left( v\right) &&\quad {m}\left( \ell ^{+}\ell ^{-}\right) \text{ }{\text{%
not between 80 and 100 GeV}\ \text{for same-flavor }\ell ^{+}\ell ^{-}.}
\end{eqnarray*}
After these cuts, less than 1 fb of $Z\rightarrow \ell ^{+}\ell ^{-}$\ ($%
\ell =e,\mu $) remains, and we do not consider it any further. The other
backgrounds, together with some representative signal points, are displayed
in Table \ref{tab:ll}. As shown in the table, the largest background after
the $Z$ veto cut $\left( v\right) $\ is $Z\rightarrow \tau ^{+}\tau ^{-}$.
Although the $Z$ in such events cannot be reconstructed in as
straightforward a manner as for $Z\rightarrow \ell ^{+}\ell ^{-}$, an
indirect method is available.

\cutgroup{ll}
\begin{table} \centering%
\begin{tabular}{|c|c|c|c|c|c|}
\hline
background & $\left( v\right) $ & $\left( vi\right) $ & $\left( vii\right) $
& $\left( viii\right) $ & $\left( ix\right) $ \\ \hline
$t\bar{t}$\quad (175) & \multicolumn{1}{|r|}{155} & \multicolumn{1}{|r|}{129}
& \multicolumn{1}{|r|}{24.7} & \multicolumn{1}{|r|}{17.0} & 
\multicolumn{1}{|r|}{10.7} \\ 
$W$ pair & \multicolumn{1}{|r|}{70} & \multicolumn{1}{|r|}{61} & 
\multicolumn{1}{|r|}{27.8} & \multicolumn{1}{|r|}{18.5} & 
\multicolumn{1}{|r|}{0.1} \\ 
$W\rightarrow e,\mu ,\tau $ & \multicolumn{1}{|r|}{21} & 
\multicolumn{1}{|r|}{18} & \multicolumn{1}{|r|}{11.1} & \multicolumn{1}{|r|}{
7.5} & \multicolumn{1}{|r|}{2.2} \\ 
$Z\rightarrow \tau ^{+}\tau ^{-}$ & \multicolumn{1}{|r|}{302} & 
\multicolumn{1}{|r|}{7} & \multicolumn{1}{|r|}{6.9} & \multicolumn{1}{|r|}{
5.7} & \multicolumn{1}{|r|}{0.2} \\ \hline
total & \multicolumn{1}{|r|}{548} & \multicolumn{1}{|r|}{215} & 
\multicolumn{1}{|r|}{70.5} & \multicolumn{1}{|r|}{48.7} & 
\multicolumn{1}{|r|}{13.2} \\ \hline
25\% & \multicolumn{1}{|r|}{137} & \multicolumn{1}{|r|}{54} & 
\multicolumn{1}{|r|}{17.6} & \multicolumn{1}{|r|}{12.1} & 
\multicolumn{1}{|r|}{3.3} \\ 
5$\sigma $ @ 2 fb$^{-1}$ & \multicolumn{1}{|r|}{83} & \multicolumn{1}{|r|}{52
} & \multicolumn{1}{|r|}{29.7} & \multicolumn{1}{|r|}{24.7} & 
\multicolumn{1}{|r|}{(3.6)} \\ \hline\hline
${\tilde{t}_{1},}$ ${\widetilde{W}_{1}}$ & $\left( v\right) $ & $\left(
vi\right) $ & $\left( vii\right) $ & $\left( viii\right) $ & $\left(
ix\right) $ \\ \hline
130, 100 & \multicolumn{1}{|r|}{48.9} & \multicolumn{1}{|r|}{41.6} & 
\multicolumn{1}{|r|}{31.7} & \multicolumn{1}{|r|}{26.8} & 
\multicolumn{1}{|r|}{13.5} \\ 
150, 110 & \multicolumn{1}{|r|}{24.9} & \multicolumn{1}{|r|}{21.1} & 
\multicolumn{1}{|r|}{14.2} & \multicolumn{1}{|r|}{12.6} & 
\multicolumn{1}{|r|}{7.2} \\ 
170, 120 & \multicolumn{1}{|r|}{12.9} & \multicolumn{1}{|r|}{10.9} & 
\multicolumn{1}{|r|}{6.4} & \multicolumn{1}{|r|}{5.6} & \multicolumn{1}{|r|}{
3.2} \\ \hline
\end{tabular}
\caption[Dilepton channel cross-sections]{%
dilepton channel cross-sections, in fb, after cuts as
discussed in the text. Signal levels required for detection are given in
the ``25\%" (of background) and ``5$\sigma$" rows.  The parenthesized $5\sigma$ value
in the last column is for an integrated luminosity of 25 $\text{fb}^{-1}$.
For the signal cases, $\tilde{t}_{1}$ and $\widetilde{W}_{1}$ masses are
given in GeV, and $m_{\widetilde{Z}_1}=m_{\widetilde{W}_1}/2$.
 \label{tab:ll}}%
\end{table}%

\miscplot{lltau}{128 228 482 641}%
{Distribution of $\mtautau$}{%
Distribution of $\mtautau$ in the dilepton channel,
after cut $(v)$.
The signal case is a 130 GeV 
${\tilde{t}_{1}}$ decaying to a 100 GeV 
${\widetilde{W}_{1}}$.
The vertical axes are in fb/bin.
Only events for which $\mtautau>0$ are shown;
the percentage of such events is indicated in parentheses
on the plots (so, 99.7\% of the \ztt\ events are plotted,
but only 41\% of the \ttbar\ events).
Dashed vertical lines indicate cut $(vi)$.
}

In the $Z$ rest frame, the $\tau $s are highly relativistic, so that the
lepton and neutrinos from $\tau \rightarrow \nu _{\tau }\nu \ell $ are
strongly boosted along the $\tau $ direction. We approximate the $\tau ^{+}$
3-momentum as $\vec{\tau}^{+}\sim P_{+}\hat{\ell}^{+}$ where $\hat{\ell}^{+}$
is the observed $\ell ^{+}$ unit direction vector and $P_{+}$ is the
(unknown) magnitude of $\vec{\tau}^{+}$. Now, the $Z$ transverse momentum
all comes from its recoil against QCD radiation, so we can estimate it as $%
\vec{Z}_{\mathrm{\top }}=-\sum_{\mathrm{hadrons}}\vec{E}_{\mathrm{\top }}(%
\mathrm{had})_{\mathstrut }^{\mathstrut }$. From above, we can also write $%
\vec{Z}_{\mathrm{\top }}=(\vec{\tau}^{+}+\vec{\tau}^{-})_{\mathrm{\top }%
}^{\mathstrut }=(P_{+}\hat{\ell}^{+}+P_{-}\hat{\ell}^{-})_{\mathrm{\top }%
}^{\mathstrut }$. Therefore after solving the two components of 
\begin{eqnarray*}
\left( P_{+}\hat{\ell}^{+}+P_{-}\hat{\ell}^{-}\right) _{\mathrm{\top }%
}^{\mathstrut }=-\sum \vec{E}_{\mathrm{\top }}(\mathrm{had}) 
\end{eqnarray*}
for the two unknowns $P_{+}$ and $P_{-}$ we can reconstruct the $\tau $
4-momenta as 
\begin{eqnarray}
\tau ^{\pm }=(\sqrt{(P_{\pm }\hat{\ell}^{\pm })^{2}+m_{\tau }^{2}},\,P_{\pm }%
\hat{\ell}^{\pm }). 
\end{eqnarray}
Note that for this procedure to make sense for $Z\rightarrow \tau ^{+}\tau
^{-}$\ both $P_{+}$ and $P_{-}$ should be positive quantities, which is to
say that neither $\tau $ should be oppositely directed to its lepton. (This
condition is also equivalent to requiring that $Z_{\mathrm{\top }}$ lie
within the smaller angle of $\ell _{\mathrm{\top }}^{+}$ and $\ell _{\mathrm{%
\top }}^{-}$.) Distributions of $m(\tau ^{+},\tau ^{-})$ after cut $\left(
v\right) $\ are displayed in Fig.~\ref{fig:lltau} for those events with $%
P_{+},P_{-}>0$. We make the cut 
\begin{eqnarray*}
\left( vi\right) \quad {m(\tau ^{+},\tau ^{-})<50\mathrm{\ GeV}\text{ or }%
m(\tau ^{+},\tau ^{-})>150\mathrm{\ GeV}} 
\end{eqnarray*}
where we define $m(\tau ^{+},\tau ^{-})=0$ if either $P_{\pm }\leq 0$.
Essentially all of the $Z\rightarrow \tau ^{+}\tau ^{-}$\ events have $%
P_{+},P_{-}>0$, while only about 40\% of the signal events do. Table \ref
{tab:ll} shows the results of applying cut $\left( vi\right) $, where we see
that only 2\% of the $Z\rightarrow \tau ^{+}\tau ^{-}$\ background survives,
while about 5/6 of the signal is retained.

\miscplot{llb}{106 104 400 588}%
{Distribution of $B$}{%
Distribution of $B$ in the dilepton channel,
after cut $(vi)$.
The signal case is a 130 GeV 
${\tilde{t}_{1}}$ decaying to a 100 GeV 
${\widetilde{Z}_{1}}$.
The vertical axes are in fb/bin.
Dashed lines at $B=120$ indicate cut $(vii)$.
}

Even after this cut, the signal is still below the $5\sigma $ level of
observability. 
To achieve a further purification we note, as in the previous channel, that
the presence in the background of two real $W$s undergoing two-body decays $%
W\rightarrow \ell \nu $ will typically result in harder lepton momenta. We
also find, for the ${\tilde{t}_{1}}$/${\widetilde{W}_{1}}$/${\widetilde{Z}%
_{1}}$ masses accessible in this channel, that the backgrounds generally
have higher $\NEG{E}\mathstrut _{\top }^{\text{ }}$. The background events
are thus ``bigger'' than the signal events, and a convenient quantity
summarizing these features is ``bigness'' 
\begin{eqnarray}
B\equiv p\mathstrut _{\top }^{\text{ }}(\ell ^{+})+p\mathstrut _{\top }^{%
\text{ }}(\ell ^{-})+\NEG{E}\mathstrut _{\top }^{\text{ }}%
\label{eq:bigness}%
\end{eqnarray}
which was first introduced in our Run I work \cite{BST}. Figure \ref{fig:llb}
shows distributions of $B$ after cut $\left( vi\right) $. We find that the
discovery reach over the parameter space is maximized with the cut 
\begin{eqnarray*}
\left( vii\right) \quad B<120\text{ GeV.} 
\end{eqnarray*}
Table \ref{tab:ll} demonstrates the utility of this cut: only 20\% of the $t%
\bar{t}$\ and 50\% of the $WW$ backgrounds survive, while we keep 60--75\%
of the signal.

\miscplot{llj}{106 104 400 588}%
{Distribution of $J_\top$}{%
Distribution of $J_\top$ in the dilepton channel,
after cut $(vii)$.
The signal case is a 130 GeV 
${\tilde{t}_{1}}$ decaying to a 100 GeV 
${\widetilde{Z}_{1}}$.
The vertical axes are in fb/bin.
Dashed lines show cut $(vii)$ as $25<J_\top<175$ GeV.
}

The reach in the dilepton channel after cut $\left( vii\right) $\ is still
substantially less than that of the $b$-jet + lepton\ channel. The situation
is improved a bit by considering the total jet transverse energy $J_{\mathrm{%
\top }}\equiv \sum_{\mathrm{jets}}\left| E_{\mathrm{\top }}^{\mathstrut
}\right| $ whose distributions are shown in Fig.~\ref{fig:llj}. $J_{\mathrm{%
\top }}$ for the $W$ and $Z$ backgrounds\ is rather lower than that of the
signal, while $J_{\mathrm{\top }}$ for $t\bar{t}$\ is higher. We use the cut 
\begin{eqnarray*}
\left( viii\right) \quad {25<J_{\mathrm{\top }}<175\mathrm{\ GeV}} 
\end{eqnarray*}
which cuts 1/3 of the background at a cost of less than 1/6 in signal, as
shown in Table \ref{tab:ll}.

\conplot{Dilepton}{
{\ Cross-section contours, in fb, for the dilepton channel after
cut }$\left( viii\right) ${\ as described in the text. The heavy solid line
at 24.7 fb is the 5$\sigma $ discovery limit for an integrated luminosity of
2 $\mathrm{fb}^{-1}$. The dashed line is 12.1 fb of signal, which is 25\% of
background. The dot-dashed line shows the reach at 25 $\mathrm{fb}^{-1}$\
after cut }$\left( ix\right) .$
}

Figure \ref{fig:ll:con} displays contours of the cross-sections after cut $%
\left( viii\right) $ in the $m_{{\tilde{t}_{1}}}$--$m_{{\widetilde{W}_{1}}}$
plane with $m_{{\widetilde{Z}_{1}}}=m_{{\widetilde{W}_{1}}}/2$ and $\mathcal{%
B}({\widetilde{W}_{1}}\rightarrow \ell )=\mathcal{B}(W\rightarrow \ell )$.
The total background is 49 fb. The $5\sigma $ discovery limit for 2 $\mathrm{%
fb}^{-1}$\ is 25 fb, shown as a heavy solid line in the figure. The dashed
line is at 12 fb = 25\% of background. The reach in this channel is
significantly less than that of the $b$-jet + lepton\ channel; ${\tilde{t}%
_{1}}$s heavier than 135 GeV are not accessible via this search.

\subsubsection{Relaxing the assumption $\mathcal{B}({\widetilde{W}_{1}}%
\rightarrow \ell )\approx \mathcal{B}({W}\rightarrow \ell )$}

Because the cross-section for dilepton events goes as the square of $%
\mathcal{B}({\widetilde{W}_{1}}\rightarrow \ell )$, substantial improvement
is possible. The situation will continue to improve as the enhancement
factor $r$ of eqn.~(\ref{eq:r}) discussed in the previous section increases
to its limit $r=3$, at which point the signal cross-section in this channel
would go up by an order of magnitude. In such a case even just 2 fb$^{-1}$
of data could find any stops with $m_{{\tilde{t}_{1}}}\lesssim m_{t}$, even
if the chargino is very close to the kinematical limit $m_{{\widetilde{W}_{1}%
}}<m_{{\tilde{t}_{1}}}+m_{b}$. Also, should a stop be discovered in the $%
\lowercase{b}%
$-jet + lepton\ channel of the previous section, then a search in the
dilepton channel might yield useful information about $\mathcal{B}({%
\widetilde{W}_{1}}\rightarrow \ell )$ which, together with basic facts about
the ${\widetilde{W}_{1}}$ discernible from the discovery, could shed some
light on other SUSY parameters.

\subsection{Run II+}

This channel is severely rate limited. We can reach the dashed 25\% contour
at 12.1 fb in Fig.~\ref{fig:ll:con} by using cuts $\left( i\right) -\left(
viii\right) $ on an 8.3 fb$^{-1}$ data sample, by eqn. (\ref{eq:400}). This
gives us stops out to $m_{{\tilde{t}_{1}}}\lesssim 150$\ GeV and charginos
to $m_{{\widetilde{W}_{1}}}\lesssim 120$ GeV. We are background limited here
and another cut is required to go further. For the 25 $\mathrm{fb}^{-1}$ Run
II+ sample, we can eliminate much of the $W$ and $Z$ background by insisting
on a tagged $b$ quark: 
\begin{eqnarray*}
\left( ix\right) \quad \text{{at least one tagged }}B.
\end{eqnarray*}
We did not impose a $b$-tag requirement for the Run II cuts, since there is
not enough signal at 2 fb$^{-1}$ to sustain such a cut. Also, Fig.~\ref
{fig:ll:con} shows that one pays a penalty for the $b$-tag, in that an
unobservable region of small $m_{{\tilde{t}_{1}}}-m_{{\widetilde{W}_{1}}}$
is created where the $b$ does not have enough energy to generate the tag.
Table \ref{tab:ll} exhibits the results of this cut. With $\left( ix\right) $%
, the 25 $\mathrm{fb}^{-1}$\ $5\sigma $ discovery limit is extended to the
dot-dashed line shown in Fig.~\ref{fig:ll:con}, which pushes the stop mass
reach to just over 165 GeV, and increases the region where the SUSY signal
can be identified in more than one channel. Detection in both channels helps
to identify that the signal is indeed a stop, and can also yield useful
information on other SUSY parameters relevant to branching fractions and
masses.

\chapter{Summary and Conclusions%
\label{chap:summ}%
}

\miscplot{summ}{115 270 475 617}%
{Summary of $5\sigma$ discovery limits for the light stop.}{%
Summary of $5\sigma$ discovery limits for the light stop.
The vertical axis on the left is neutralino mass, and that on the right
is chargino mass.  In case $m_{\widetilde{W}_1}=2m_{\widetilde{Z}_1}$,
these axes coincide as illustrated.
Solid contours show the reach at 2 fb$\mathstrut^{-1}$; dashed lines are for
 25 fb$\mathstrut^{-1}$.
The irregularly shaped hatched region at the left is excluded at the 95\%
confidence level in the
$\scz$ mode by LEP II and Tevatron Run I experiments.
The hatched region under $m_\wino=93$ is excluded by LEP II
chargino mass limit.
}
In this work we have studied two important stop decay modes, ${\tilde{t}_{1}}%
\rightarrow c{\widetilde{Z}_{1}}$ and ${\tilde{t}_{1}}\rightarrow b{%
\widetilde{W}_{1}}$, in the context of the Run II Tevatron experiments. We
have identified useful kinematical quantities for distinguishing the stop
signal events from SM backgrounds. We found that these backgrounds could be
controlled well enough to expose the stop over a significant range of the
parameter space of many models, so that Tevatron experiments in the Main
Injector era will probe models not accessible to LEP\ II.

Our results are summarized in Fig.~\ref{fig:summ}. Both the $m_{{\tilde{t}%
_{1}}}\times m_{{\widetilde{Z}_{1}}}$ and $m_{{\tilde{t}_{1}}}\times m_{{%
\widetilde{W}_{1}}}$ planes are displayed, after the fashion of Fig.~\ref
{fig:expt}. As with that figure, the vertical axes for $m_{{\widetilde{Z}_{1}%
}}$ and $m_{{\widetilde{W}_{1}}}$ will coincide in case $m_{{\widetilde{W}%
_{1}}}=2m_{{\widetilde{Z}_{1}}}$\ (indeed, the ${\tilde{t}_{1}}\rightarrow b{%
\widetilde{W}_{1}}$ contours correspond to $m_{{\widetilde{W}_{1}}}=2m_{{%
\widetilde{Z}_{1}}}$).

For the ${\tilde{t}_{1}}\rightarrow c{\widetilde{Z}_{1}}$ mode we show the
discovery limits at 2 fb$^{-1}$ and 25 fb$^{-1}$, for both the $\NEG%
{E}\mathstrut _{\top }^{\text{ }}$ channel and the $c$-tagged channel. The
irregular hatched region is excluded at the 95\% confidence level by LEP\ II
and Tevatron Run I experiments. Figure \ref{fig:summ} shows the $c$-tagged
channel channel to be clearly superior to the $\NEG{E}\mathstrut _{\top }^{%
\text{ }}$ channel\ for this decay mode. While the reach in ${\tilde{t}_{1}}$
mass is comparable for the two, $c$-tagging allows heavier $m_{{\widetilde{Z}%
_{1}}}$ to be probed. This is chiefly due to the large jet $E_{\top }$
requirement in the $\NEG{E}\mathstrut _{\top }^{\text{ }}+$ jets search
which means that $m_{{\tilde{t}_{1}}}-m_{{\widetilde{Z}_{1}}}$ must be large
to yield hard $c$-jets.\ In either channel, we may find stops with masses
below 165 GeV in Run II. The $c$-tagged channel generally gives an extra 15
GeV of reach in ${\widetilde{Z}_{1}}$ mass.

For higher integrated luminosities, the stop mass may be pushed well past
the top mass, and $c$-tagging with 25 fb$^{-1}$ of data allows discovery of
stops heavier than 200 GeV. In the Run II+ scenario, $c$-tagging extends the 
${\widetilde{Z}_{1}}$ mass reach by 20 to 30 GeV over the $\NEG{E}\mathstrut
_{\top }^{\text{ }}$ channel. Note however that as the ${\tilde{t}_{1}}$
becomes heavier the four-body decay discussed in Section \ref{sec:stop:prod}
may have a substantial branching fraction.

Results for the ${\tilde{t}_{1}}\rightarrow b{\widetilde{W}_{1}}$ decay mode
are shown in the lower right portion of Fig.~\ref{fig:summ}. The $\ell +b$%
-jet and dilepton channels are illustrated, under our usual assumptions $m_{{%
\widetilde{W}_{1}}}=2m_{{\widetilde{Z}_{1}}}$ and $\mathcal{B}({\widetilde{W}%
_{1}}\rightarrow $ leptons$)=\mathcal{B}({W}\rightarrow $ leptons$)$. The
hatched region $m_{{\widetilde{W}_{1}}}<93$ GeV has been excluded by LEP\ II
at the 95\% confidence level.

For this mode, with $W$-like chargino branching fractions, the dilepton
signal is of scant use compared to the $\ell +b$-jet channel with just 2 fb$%
^{-1}$. However, when $\mathcal{B}({\widetilde{W}_{1}}\rightarrow $ leptons$%
)\gg \mathcal{B}({W}\rightarrow $ leptons$)$ the dilepton channel rapidly
improves. A detection in either or both of these channels could, by
differential analysis of the two, yield valuable information about other
sectors of the MSSM, particularly if kinematical evidence could be used to
constrain the SUSY masses. Even supposing $\mathcal{B}({\widetilde{W}_{1}}%
\rightarrow $ leptons$)=\mathcal{B}({W}\rightarrow $ leptons$)$, we could
probe stops out past 165 GeV with the data of Run II, and push this another
20 GeV or so with the higher integrated luminosity of Run II+.

Note that the Run II+ discovery frontiers plotted in Fig.~\ref{fig:summ} for
the $\ell +b$ and dilepton channels coincide for $m_{{\tilde{t}_{1}}}<165$
GeV at the high $m_{{\widetilde{W}_{1}}}$ end of the range. This reflects
the fact that the $\ell +b$ signal is strongly attenuated in the limit $m_{{%
\tilde{t}_{1}}}\rightarrow m_{{\widetilde{W}_{1}}}$ as the available energy
for a hard identifiable $b$-jet decreases, while the dilepton suffers less
in this limit. Since the statistical significance of the two channels is
comparable in this region, the two could be statistically combined to gain a
bit more reach in $m_{{\widetilde{W}_{1}}}$.

In conclusion, we have demonstrated that the Fermilab Tevatron can probe
significant and interesting portions of the light stop parameter space with
experiments conducted during the initial and extended Run II operating
phases. Particularly important, the procedures we have developed here should
allow the Tevatron to test most of the region $m_{{\tilde{t}_{1}}}\lesssim
165$ GeV favored by electroweak baryogenesis within mSUGRA models with
minimal low-energy particle content.

The large top Yukawa coupling and the availability of substantial LR mixing
due to the trilinear scalar coupling make the stop the lightest sfermion in
many models, and it may, in principle, be the lightest charged
supersymmetric particle. The fact that the Tevatron can subject this
hypothesis to a strong test holds out the exciting possibility that
supersymmetry may make its first appearance in the discovery of the stop.

%
\def\PhysRev#1#2{Phys.\ Rev.\ {\bf #1}, #2}
\def\PhysRevLett#1#2{Phys.\ Rev.\ Lett.\ {\bf #1}, #2}
\def\NuclPhysB#1#2{Nucl.\ Phys.\ {\bf B#1}, #2}
\def\PhysLett#1#2{Phys.\ Lett.\ {\bf #1}, #2}
\def\PhysRep#1#2{Phys.\ Rept.\ {\bf #1}, #2}
\def\IJMP#1#2{Int.\ J.\ Mod.\ Phys.\ {\bf #1}, #2}
\def\hepph#1{hep-ph/{\bf #1}}
\def\hepex#1{hep-ph/{\bf #1}}
\def\ModPhysLett#1#2{Mod.\ Phys.\ Lett.\ {\bf #1}, #2}

\end{document}